\documentclass[sigplan,10pt]{acmart}
\acmSubmissionID{<PAPER ID>}
\renewcommand\footnotetextcopyrightpermission[1]{} 

\acmDOI{00.0000/0000000.0000000}

\usepackage{hyperref}
\usepackage{multirow}

\usepackage{filecontents}

\usepackage{graphicx}
\usepackage{tikz}
\usepackage{caption}
\usepackage{subfig}

\usepackage{circledsteps}
\usepackage{xcolor,colortbl}
\usepackage{multirow}
\usepackage{tabularx}
\usepackage{cleveref}
\usepackage{tcolorbox}
\usepackage{float}
\usepackage{balance}

\usepackage{subfloat}
\definecolor{customgrey}{RGB}{210,210,210}

\begin{document}

\setcopyright{none}
\acmConference[EuroSys '25]{Twentieth European Conference on Computer Systems}{March 30--April 3, 2025}{Rotterdam, Netherlands (preprint)}


\title[Serverless Cold Starts and Where to Find Them]{Serverless Cold Starts and Where to Find Them}

\author{Artjom Joosen, Ahmed~Hassan, Martin Asenov, \\Rajkarn Singh, Luke~Darlow, Jianfeng Wang, Qiwen Deng, Adam Barker}
\email{sirlab@huawei.com}
\affiliation{
  \institution{Central Software Institute, Huawei}
  \country{}
}

\newcommand{\revision}[1]{#1} 

\renewcommand{\shortauthors}{Joosen, et al.}

\begin{abstract}
This paper analyzes a month-long trace of 85 billion user requests and 11.9 million cold starts from Huawei's serverless cloud platform. Our analysis spans workloads from five data centers. We focus on cold starts and provide a comprehensive examination of the underlying factors influencing the number and duration of cold starts. These factors include trigger types, request synchronicity, runtime languages, and function resource allocations. We investigate components of cold starts, including pod allocation time, code and dependency deployment time, and scheduling delays, and examine their relationships with runtime languages, trigger types, and resource allocation. We introduce pod utility ratio to measure the pod's useful lifetime relative to its cold start time, giving a more complete picture of cold starts, and see that some pods with long cold start times have longer useful lifetimes. Our findings reveal the complexity and multifaceted origins of the number, duration, and characteristics of cold starts, driven by differences in trigger types, runtime languages, and function resource allocations. For example, cold starts in Region 1 take up to 7 seconds, dominated by dependency deployment time and scheduling. In Region 2, cold starts take up to 3 seconds and are dominated by pod allocation time. Based on this, we identify opportunities to reduce the number and duration of cold starts using strategies for multi-region scheduling. Finally, we suggest directions for future research to address these challenges and enhance the performance of serverless cloud platforms. 

\end{abstract}



\maketitle

\section{Introduction}

Serverless computing has become a widely used computing paradigm, with all major cloud providers expanding and improving their serverless offerings. Serverless platforms rely on systems such as Knative~\cite{Knative}, YuanRong~\cite{chen2024yuanrong}, and Nuclio~\cite{Nuclio} running on top of container management systems such as Kubernetes~\cite{li2021analyzing} to manage the infrastructure.  This allows developers to focus on application code, written as functions, while the cloud service provider manages the underlying infrastructure. 

For each function, users specify triggers, such as timers, that activate the function. When a function is triggered, a container with that function's code processes the request. If a new container needs to be started either because there are no active containers for this function or because existing containers are overloaded, this causes a \emph{cold start}. This cold start adds significant latency, degrading application performance. Hence, there has been considerable effort to optimize serverless applications~\cite{mahgoub2021sonic,li2023modernization}, frameworks~\cite{agache2020firecracker, zhuang2023exoflow}, and their resource usage~\cite{zhang2021faster,liu2023doing}. 


Cold starts are a significant challenge in serverless systems~\cite{mahgoub2022orion, li2022help, chen2022starlight,wei2023no}. While an important problem, there is a lack of open-source data on cold starts from production systems. Currently, all available data from production serverless systems contain only aggregate statistics~\cite{shahrad2020serverless,Joosen_2023}.
While it has been suggested that cold starts are affected by trigger types, package sizes, and language runtimes, among other factors~\cite{Joosen_2023}, to the best of our knowledge, there exists no detailed analysis of the relationship between these factors and cold starts in production deployments. To solve this problem, some studies have tried to measure cold starts from a user perspective and relate them to their sources~\cite{wang2018}. However, these measurements may not be representative of workloads run by real users. Additionally, previous research lacks the provider's view of long-term trends across multiple data centers, making it difficult to suggest optimizations that exploit workload patterns observed in production workloads. 

In this paper, we present a thorough analysis of 31 days of detailed metrics from serverless deployments in five of Huawei Cloud's data center regions. The data includes 85 billion requests from over 12 million pods, including over 11 million cold starts\footnote{Huawei traces available at \url{https://github.com/sir-lab/data-release}}. Our analysis highlights differences in peak time effects, resource usage, and cold start distributions between and within our data centers. Furthermore, we analyze the distribution and number of cold starts for different runtimes, trigger types, and resource allocations to identify which workloads tend to cause cold starts. Finally, the period studied in this paper contains a week-long holiday, enabling us to analyze workload changes before, during, and after a holiday. 

This paper provides the results of our in-depth analysis as primary contributions. Specifically, we:
\begin{enumerate}
    \item Analyze cold starts in production serverless deployments, including how cold starts relate to trigger types, synchronicity, language runtimes, and pod sizes.
    \item Characterize factors that impact how cold starts differ across five different data center regions, providing insights into how different regions within the same cloud provider can exhibit varying performance.
    \item Provide a first-of-its-kind component analysis on the different latency components in cold starts, providing insights into the causes of cold starts, and insights about where and how cold starts should be optimized.
    \item  We study long-term time-varying effects of cold starts and components, and identify the potential for spatial peak shaving between regions due to differing peak times and temporal peak shaving, which is enabled by substantial delayed pod allocations.
    \item  Based on our insights, we identify areas for improvement in data center scheduling and design, and provide several directions for future research, including resource pool prediction, concurrency adjustment and call chain prediction. 
\end{enumerate}

\section{Background}

Serverless computing allows developers to deploy event-driven functions without provisioning or managing servers or backend infrastructure~\cite{agache2020firecracker}. This section starts by describing our platform and dataset. We then provide a brief background on cold starts, followed by a description of the analysis we perform on the dataset, and the shortcomings of recent works analysing serverless workloads.

\subsection{Our Platform and Dataset}
Huawei's serverless offerings rely on an in-house platform, YuanRong, a general-purpose serverless platform with a unified programming interface. YuanRong has been deployed for over three years across nearly 20 datacenter regions, processing up to 30 billion requests per day~\cite{chen2024yuanrong}. Each region is divided into four clusters. Clusters provide virtual and physical separations within a region, improving availability and fault tolerance. 

Huawei's public cloud lets users upload function code and assign a resource limit for the function. These resource limits are grouped into CPU-memory configurations, such as `300-128', representing 300 millicores and 128MB of memory. Pools of pods with different resource configurations are maintained in case one is required by a cold start. If the autoscaler determines that additional pods are required to address incoming requests, pods are taken from the appropriate pool, the code of that function is loaded into it, and it is ready to process requests. 

Requests for a given function may be balanced between clusters or routed to a single cluster. Several load-balancers receive requests and dispatch them to the nodes running function instances. The load-balancers keep track of the number of requests dispatched but not yet returned for each function. If there is a certain cluster with increased load, the system will balance traffic between the clusters within the region, starting pods in a new cluster if they do not exist, or shifting the load to existing pods. If there are no hot-spots in the clusters, a hash mechanism is used to dispatch requests to only one cluster.

\begin{small}
\begin{table}
\renewcommand{\arraystretch}{0.9}
	\centering
    \begin{tabular}{ |c|c|c| } 
            \hline
            \multicolumn{3}{|c|}{\textbf{Request level table, Regions: 5, Duration: 31 days}}\\
            \hline
            Name & Description & Res  \\
            \hline
            Timestamp & timestamp at worker & ms \\
            Pod ID & hashed pod ID & -  \\
            Cluster name & cluster name & -  \\
            Function name & hashed function name & -  \\
            User ID & hashed user ID & - \\
            Request ID & hashed request ID & -  \\
            Execution time & execution time & $\mu$s \\
            CPU usage & CPU usage & millicores  \\
            Memory usage & memory usage & Bytes \\
            \hline
            \multicolumn{3}{|c|}{\textbf{Pod level table, Regions: 5, Duration: 31 days}}\\
            \hline
            Timestamp & timestamp & ms \\
            Pod ID & hashed pod ID & -  \\
            Cluster name & cluster name & -  \\
            Function name & hashed function name & -  \\
            User ID & hashed user ID & -  \\
            Cold start time & total cold start time & $\mu$s \\
            Pod alloc. time & time to get pod from pool & $\mu$s  \\
            Deploy code time & time to deploy code & $\mu$s  \\
            Deploy dep. time & deploy dependency time & $\mu$s \\
            Scheduling time & scheduling overhead time & $\mu$s \\
            \hline
            \multicolumn{3}{|c|}{\textbf{Function level table, Regions: 1, Duration: 31 days}}\\
            \hline
            Function name & hashed function name & -  \\
            Runtime & runtime & - \\
            Trigger type & trigger type & -  \\
            CPU-MEM & CPU-MEM config & -  \\
            \hline
    \end{tabular}
	\caption{Summary of our dataset fields with each field's associated timestamp granularity. }
 \label{tab:datasets}
\end{table}
\end{small}

\paragraph{Our dataset.} In this paper, we analyze detailed metrics from five different regions collected from a total of 20 clusters. The data does not include all functions from all regions in our data centers, but provides a good representation of how serverless production systems operate across multiple regions. The dataset comes from three different monitoring streams: request level monitoring, pod level monitoring, and function level monitoring. From the request level monitoring we analyze and release data from five regions including: timestamp, pod ID, function name, user ID of the function owner, request ID, request execution time, as well as CPU and memory usage of the request. From pod level monitoring, we analyse data logged during cold start events, including timestamp, pod ID, cluster ID hosting the pod, function name, user ID, total cold start time, and components of different parts of the cold start including pod allocation time, code deployment time, and dependency deployment time. We also analyze trigger types of different functions along with their CPU and memory configuration, and the runtime of these functions. For privacy reasons, all IDs are hashed. We summarize the fields we use in our analysis in Table~\ref{tab:datasets}.

\begin{figure}[h]
	\centering
        \includegraphics[width=0.99\linewidth]{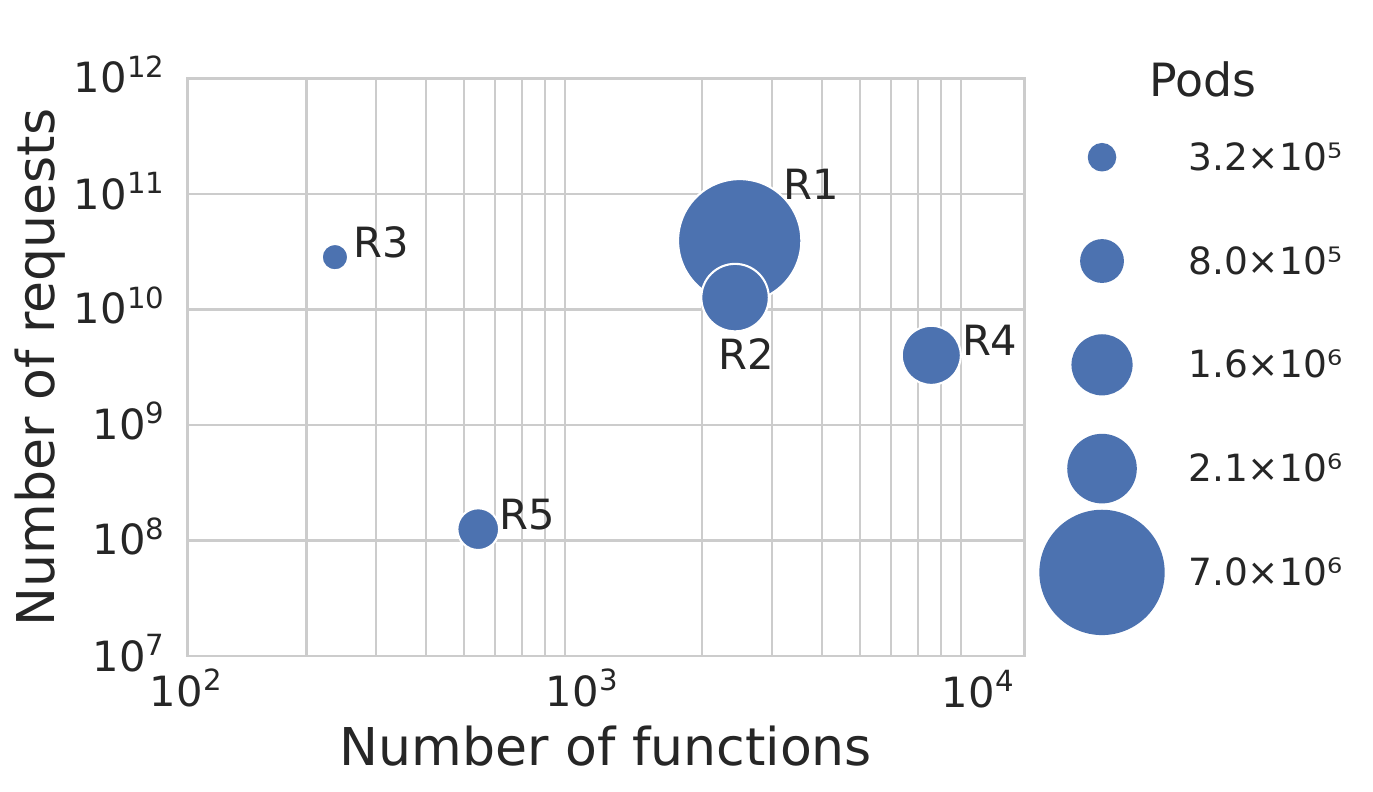}  
	\caption{Plot showing the number of requests, functions, and pods for all five regions.}
    \label{fig:plot_unique_functions}
\end{figure}

To the best of our knowledge, this is the most detailed serverless dataset released by any cloud provider. The dataset has event-level metrics for a total of 85 billion requests and over 11 million cold starts over a period of 31 days with a week-long major holiday period. Figure ~\ref{fig:plot_unique_functions} gives an overview of the size of the data for each region, numbered R1, R2, R3, R4, and R5. Most functions run on multiple clusters in a region. Finally, some functions have no load-balancing in a single region and are deployed only on a single cluster. The dataset hence aims to capture most of the possible dynamics for serverless production systems.

\subsection{Cold starts in our System}\label{section:cold_starts_background}

\begin{figure}[h]
	\centering
    \includegraphics[width=1\linewidth]{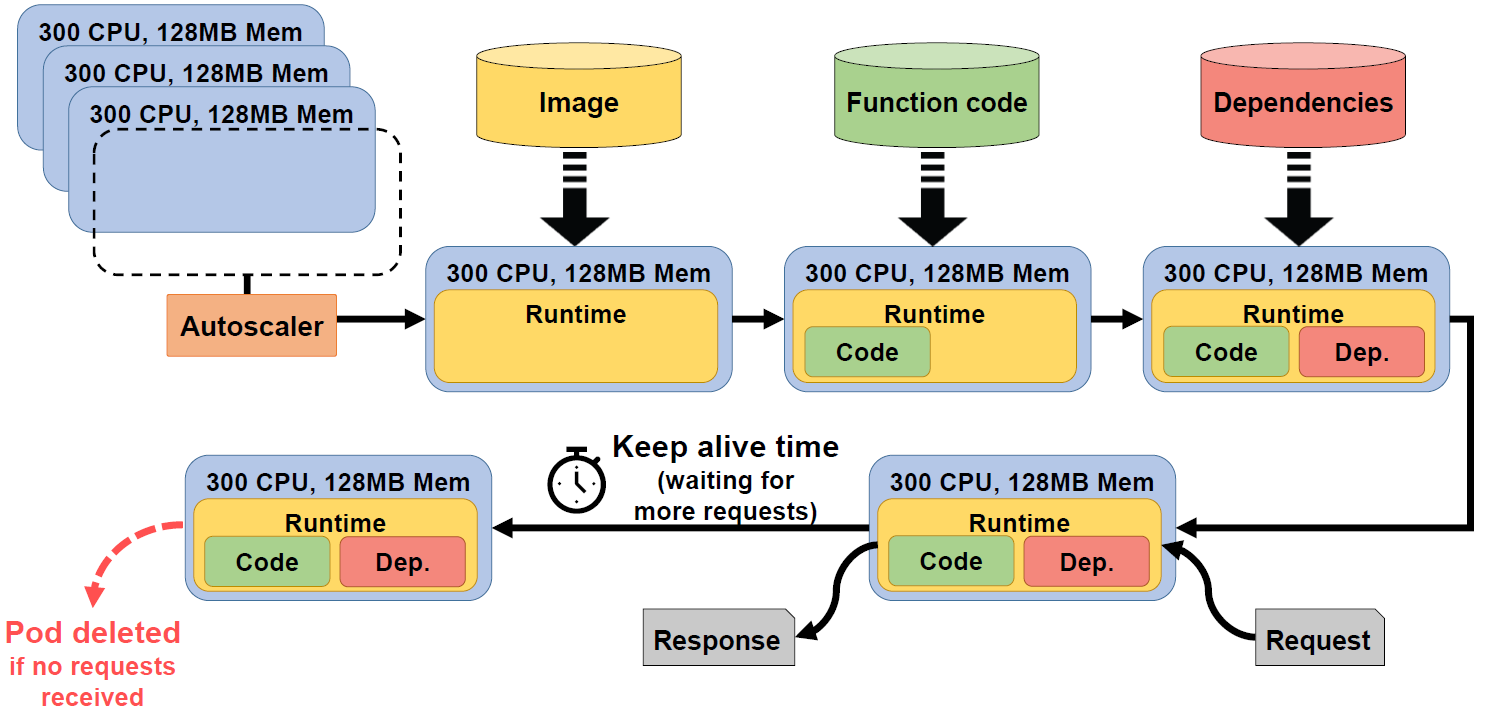}  
	\caption{Life cycle of a pod. The pod is taken from its resource pool, loaded with a runtime, function code, and dependencies. After serving requests, the pod waits for additional requests for a designated keep alive time. If the pod receives no more requests during this time, it is deleted.}
    \label{fig:pod_life_cycle}
\end{figure}

One of the main bottlenecks of serverless systems is cold starts, where the system must start a new pod if it does not have sufficient resources to process an incoming request~\cite{Joosen_2023,li2022help,wang2018}. The number and time of cold starts vary widely across different runtimes, trigger types, and resource allocations. We analyze how cold starts are affected by these factors. In addition, we analyze the times spent in the different parts of our serverless system during a cold start, as shown in Figure \ref{fig:pod_life_cycle}, including pod allocation, deploying the compressed function file, deploying dependencies, and scheduling.

As shown in Figure \ref{fig:pod_life_cycle}, our system maintains resource pools of inactive pods with different CPU and memory allocations, e.g. 300 millicores with 128MB of memory. When a function is cold-started, the scheduler selects a pod from the pool with that CPU and memory configuration. Pods have several popular runtimes and libraries preinstalled. Users can also provide a custom container image for other runtimes not supported by default, which is downloaded into the pod when needed. When the runtime is ready, the function code is pulled. Additional dependencies are downloaded if required. Once the pod is prepared, it serves the request that spawned it and returns the result. The pod then waits for additional requests for a designated keep alive time. In our system, this time is set to one minute by default. If the pod receives no requests during this time, it is deleted. Otherwise, the keep-alive time is reset back to one minute with each new request.

\section{Multi-region Serverless in Production}\label{section:serverless_in_production}

In this section, we provide a high-level overview of similarities and differences between our regions. Figure \ref{fig:plot_unique_functions} shows the size of each region by their number of requests, functions, and pods, showing differences of several orders of magnitude between regions. We observe that a larger number of functions does not necessarily mean more requests or pods. We start our analysis in this section by examining differences between regions in their resource usage, latency, peak times, peak-to-trough ratio, and holiday effects.

\subsection{Region Statistics}

\begin{figure*}
\centering
        \begin{center}
            \subfloat[CDF of sum of requests on median day per \\function for each region.]{
                \includegraphics[height=3.69cm]{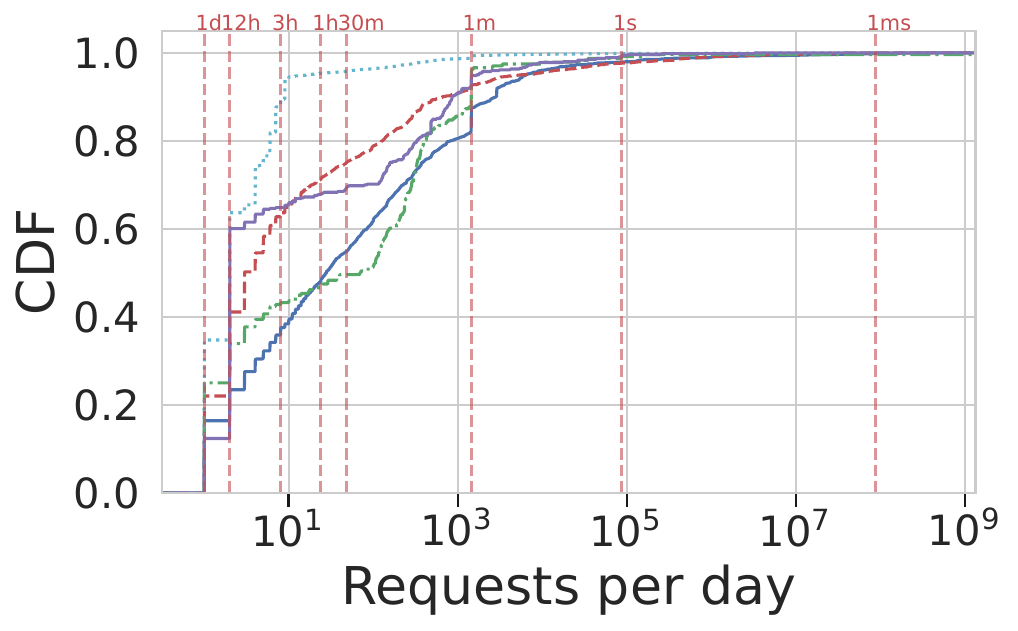}
                \label{fig:requests_cdf}}
            \hfill
            \subfloat[CDF of mean execution times per \\minute for each region.]{
              \includegraphics[height=3.57cm]{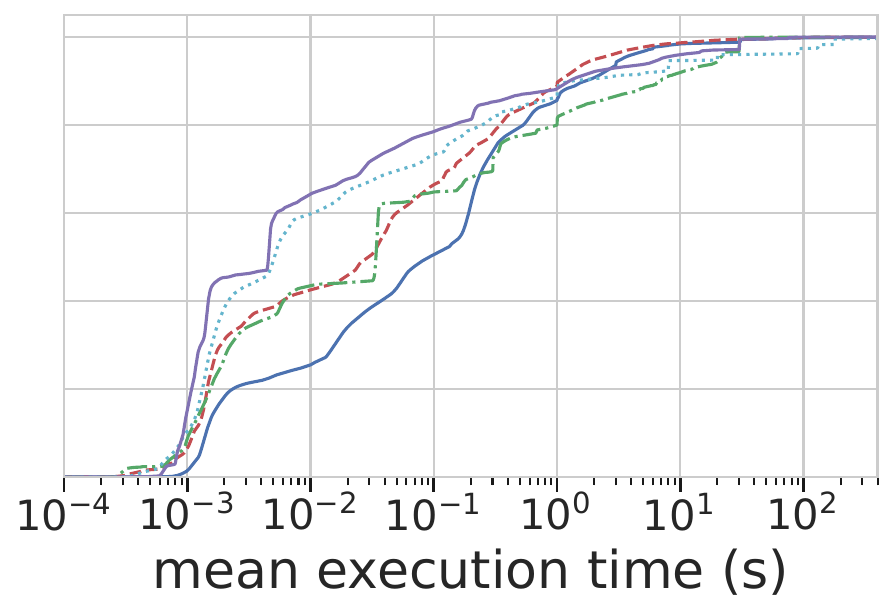}
                \label{fig:totalCost_cdf}}
            \hfill
            \subfloat[CDF of mean CPU usage in cores per minute \\for each region.]{
              \includegraphics[height=3.57cm]{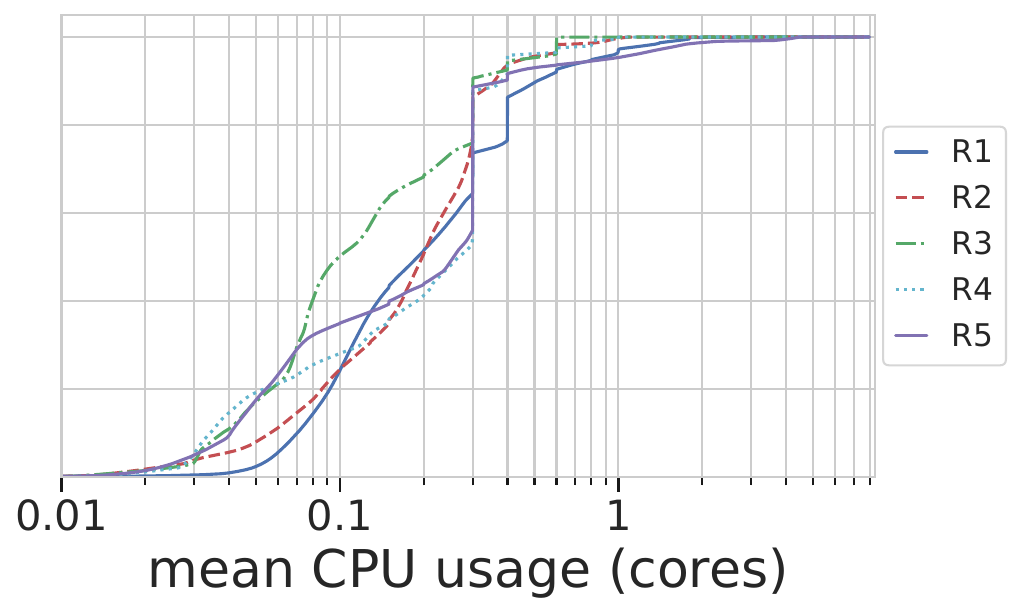}
                \label{fig:cpu_usage_avg}}
            \caption{CDF of invocations, execution time, and CPU usage for all regions.}
            \label{fig:cdfs_summary}
        \end{center}
\end{figure*}

Figures~\ref{fig:requests_cdf}, shows CDFs of the number of requests per function per day in each region, with dashed vertical lines representing the equivalent mean request inter-arrival time per function. A large majority of functions have very few requests per day on average. However, a small minority of functions have a very large number of requests and dominate resource usage. There are also large differences between regions in the number of functions with high load. Approximately 20\% of functions in Region 1 have at least one request per minute on average, while this is approximately 1\% for Region 4. We note that the average hides the per function patterns. Many of our functions have a large variation in their inter-arrival rates throughout the day.
 
Figures~\ref{fig:totalCost_cdf}, and~\ref{fig:cpu_usage_avg} show the mean execution time per minute and mean CPU usage per minute. Similar to the variations in number of arrivals per function, we also see significant variation in execution times and the CPU usage per region. Median execution time varies between 4ms in Region 5 to 100ms in Region 1. Execution time can be as long as several tens or hundreds of seconds. Additionally, the median CPU usage varies between 0.1 cores in Region 3 to 0.3 cores in Region 3. These significant differences in CPU usage and execution time between regions present opportunities for cross-region scheduling.

\begin{figure}[h]
        \begin{center}
            \subfloat[Functions per user for all regions.]{
                \includegraphics[height=3.2cm]{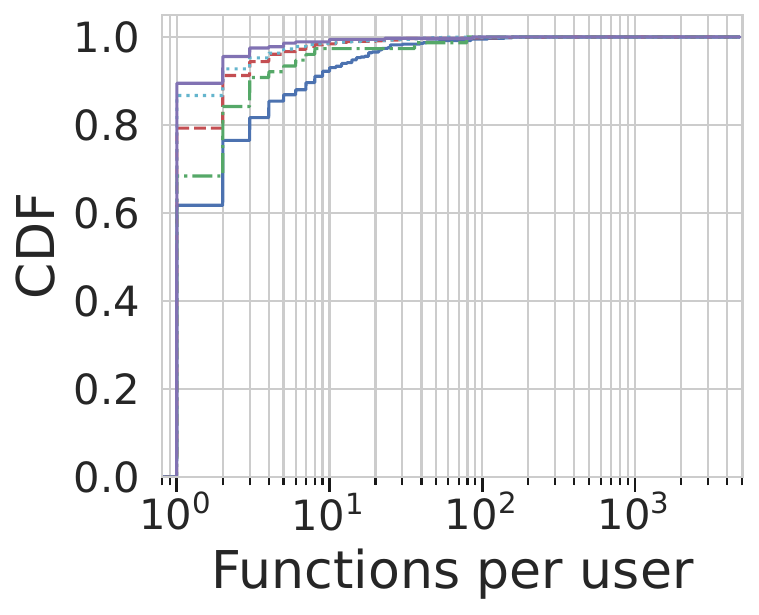}
                \label{fig:funcName_per_userID}}
            \hfill
            \subfloat[Requests per user for all regions.]{
              \includegraphics[height=3.2cm]{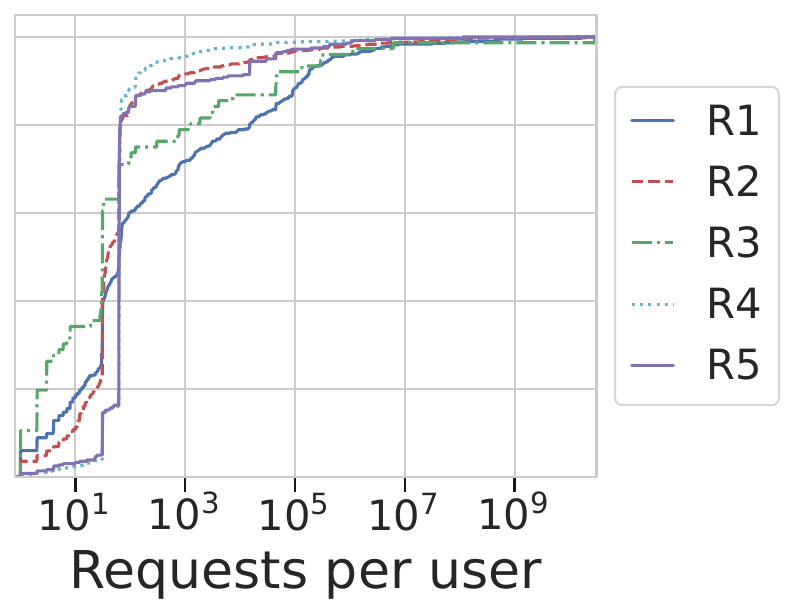}
                \label{fig:num_requests_per_userID}}
            \caption{Number of functions per user and number of requests per user for all regions over the duration of the trace.}
            \label{fig:CDFs_user}
        \end{center}
\end{figure}

\paragraph{Functions per user.} Figure \ref{fig:funcName_per_userID} shows the CDF of number of functions per user while Figure \ref{fig:num_requests_per_userID} shows number of requests per user in each region. Figure \ref{fig:funcName_per_userID} shows that 60\% to 90\% of users have only a single function, depending on the region, with almost all users having fewer than 20 functions. The two figures show that the number of functions and requests tend to be more concentrated in fewer users in smaller regions, and less concentrated in larger regions. For example, 30\% of users in Region 1 have more than 1000 requests, while less than 5\% of users have 1000 requests in Region 4. This means that, for many users, migrating their functions to a different region may be feasible due to a small number of functions.

Previous work has shown that users understand their own resource requirements poorly, and tend to choose regions close to their end users, not considering cheaper or faster options~\cite{shi2022characterizing}. This is partly because users do not know the behaviours of other users, unlike the cloud provider, which can see user behaviours and schedule more efficiently than users can. 
In addition, our regions are spread out geographically, with power prices and carbon costs varying accordingly. As we show later, the latency between regions can be insignificant compared to the longer cold starts and execution times in the more popular regions. By moving some requests to other regions, it may be possible to improve latency, cost, or carbon footprint~\cite{anderson2023treehouse, bashir2021enabling}. Such approaches have been previously suggested for batch workloads~\cite{hanafy2023carbonscaler}. However, we believe that it can also be used for serverless systems with minimal effect on execution time.

\begin{tcolorbox}[title={\centering Cross-region scheduling potential},colframe=customgrey, coltitle=black]
Regions can have very different profiles. For example, median invocations per function, function execution time, and CPU usage between regions vary by factors of up to 50, 25, and 3, respectively, which presents opportunities for cross-region load balancing. Enabling such scheduling has the potential to reduce latency, cold starts, and costs in serverless public clouds.
\end{tcolorbox}

\subsection{Peak Time Analysis}


Peak times are a well-known phenomenon in serverless systems, with several earlier works concerned with predicting and mitigating such surges~\cite{Joosen_2023, shahrad2020serverless, ResourceCentral2017}. Peak times present challenges such as large fluctuations in resource allocation and network congestion, causing increased latency. However, to the best of our knowledge, peak workloads have not been studied across multiple regions. In this section, we examine the changes that occur in the number of requests during peaks and troughs. Other prior work has shown that different peak times present opportunities for peak load shaving and can be exploited for scheduling and load balancing, as has been done in some private serverless cloud platforms~\cite{XFaaS_meta_sosp_23}. Such solutions are particularly viable in scenarios consisting of many non-latency critical workloads. 

\begin{figure}[h]
	\centering
    \includegraphics[width=0.99\linewidth]{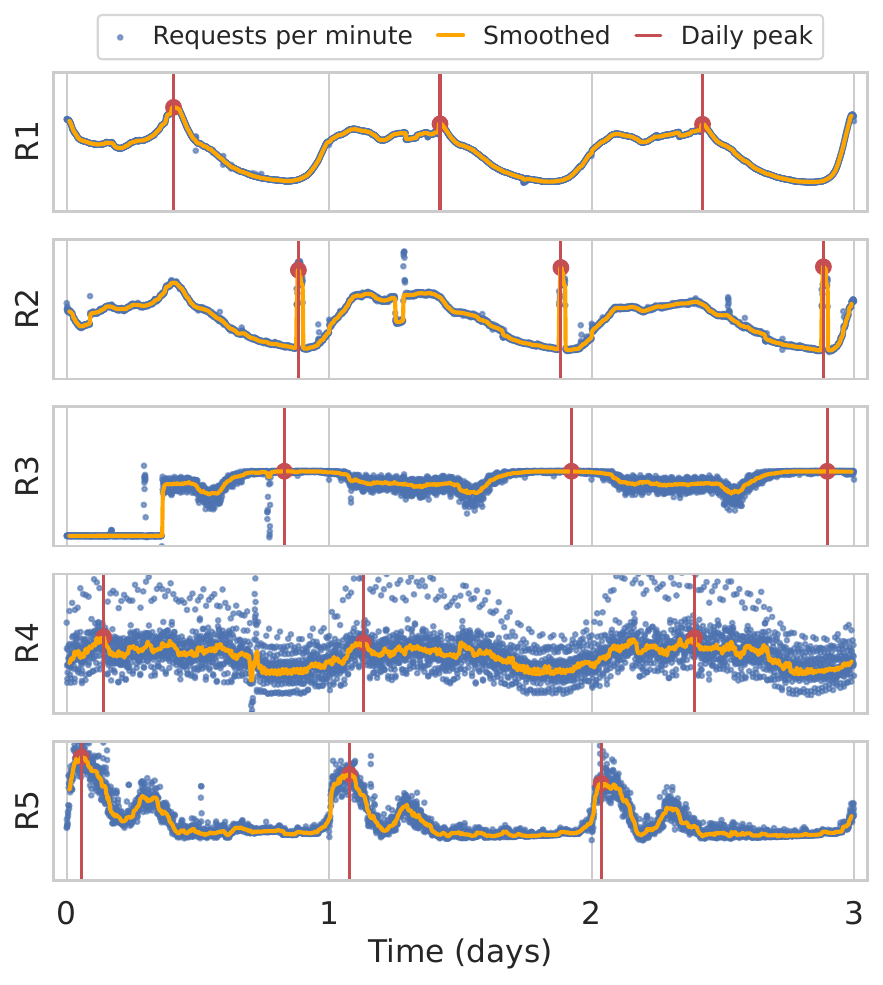}  
	\caption{Plot showing peaks in normalized number of requests per region. Peaks detected on a smoothed version of the signal. The largest peak in 24 hours is highlighted.}
    \label{fig:timeseries_sum_requests}
\end{figure}

\paragraph{Peak time lags.} Peak times occur in each one of our regions. However, they often occur at different times of the day. Figure~\ref{fig:timeseries_sum_requests} shows the normalized request patterns for each of our regions for a three day period. The largest peak every day is marked with a red line. We see clear periodic behaviour in all of our regions, consistent with existing literature~\cite{Joosen_2023, shahrad2020serverless}. The largest peaks tend to occur at a different time of day in every region, with smaller secondary peaks also present in many regions. Some regions have more prominent periodic behavior compared to others.

\paragraph{Peak-to-trough ratios for function invocations.} One way of analyzing workload periodicity and burstiness is to measure their peak-to-trough ratios, which is the ratio of the largest peak in a periodic pattern to its lowest trough. This is a measure of the strength of periodic oscillation. Large peak-to-trough ratios indicate a function with large bursts of invocations. 

\begin{figure}[h]
        \begin{center}
            \subfloat[Scatter plot of functions by their median requests per day and peak-to-trough ratio.]{
                \includegraphics[height=3.6cm]{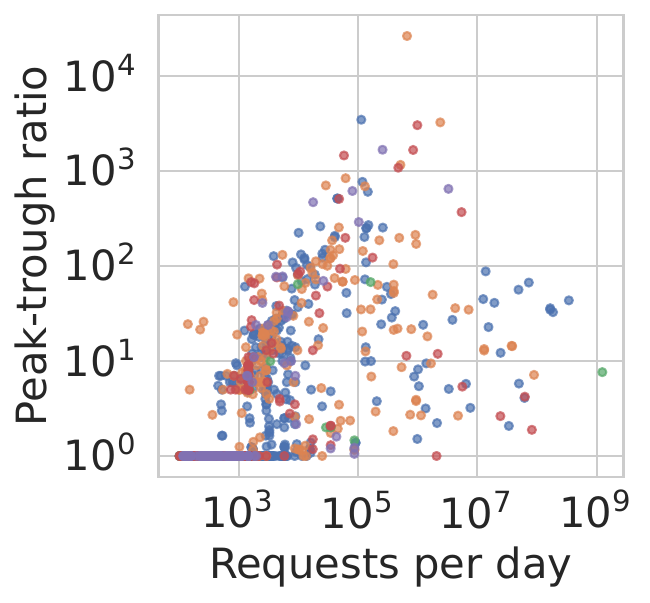}
                \label{fig:peak_trough_ratio_vs_sum_median_avg_day_all_regions}}
            \hfill
            \subfloat[Scatter plot of number of cold starts over 31 days for a function against its peak-to-trough ratio.]{
              \includegraphics[height=3.6cm]{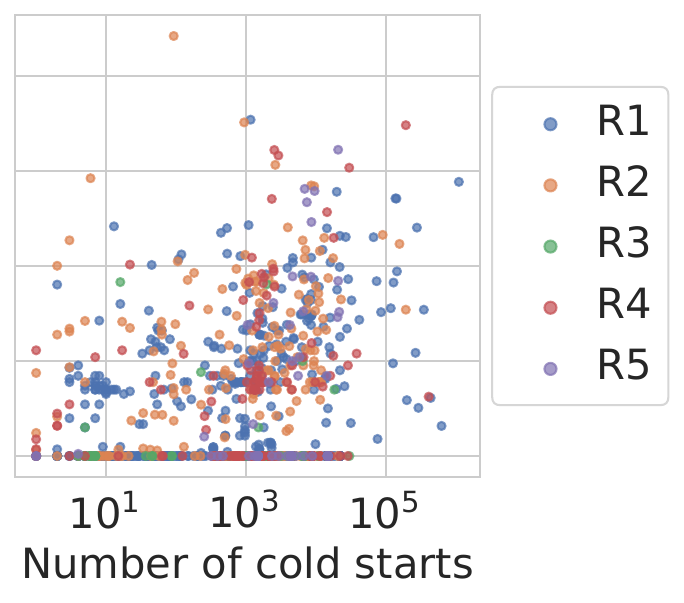}
              \label{fig:peak_trough_ratio_vs_cold_starts_all_regions}}
            \caption{Scatter plots showing a function's requests per day and number of cold starts against its peak to trough ratio. Functions with a constant value of requests per minute, or no identifiable peaks have a peak to trough ratio of one.}
            \label{fig:peak_trough_ratio}
        \end{center}
\end{figure}

Figure \ref{fig:peak_trough_ratio_vs_sum_median_avg_day_all_regions} shows a scatter plot of each function's peak-to-trough ratio against its median invocations per day. Peak-to-trough ratios tend to be lower for low-request functions, high for moderately popular functions, and lower again for very popular functions. We see a wide variety of peak-to-trough ratios, from less than 2 to over 1000. Huge variations are less common in workloads with larger number of requests, and the largest workloads experience peak-to-trough ratios less than 60. Additionally, we see a cluster of functions with fewer than 1440 requests per day (corresponding to 1 request per minute) with peak-to-trough ratios close or equal to 1. These functions are invoked on average once per minute and do not have enough requests to have identifiable peaks. Other works have previously reported peak-to-trough ratios of similar magnitude in production serverless public clouds (over 500)~\cite{FaaSNet_2021}.

\begin{figure}[h]
        \begin{center}
            \subfloat[Change in allocated pods.]{
              \includegraphics[height=3.8cm]{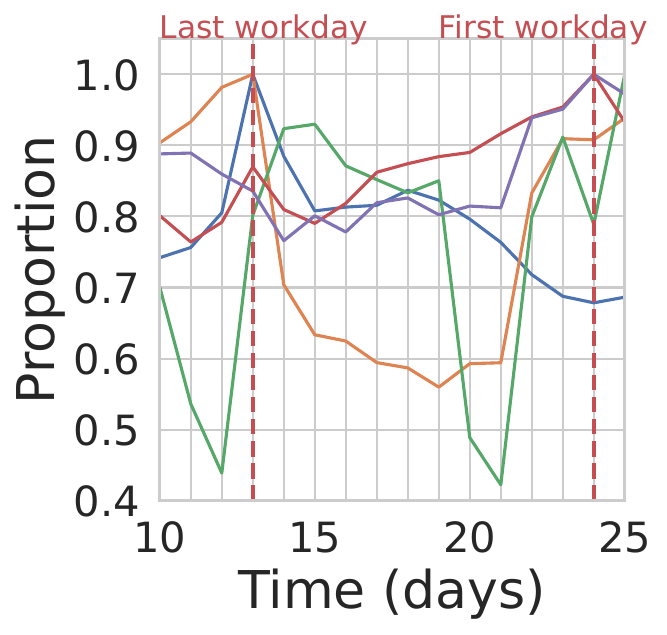}
                \label{fig:percent_change_num_pods}}
            \subfloat[Change in mean CPU usage.]{
                \includegraphics[height=3.8cm]{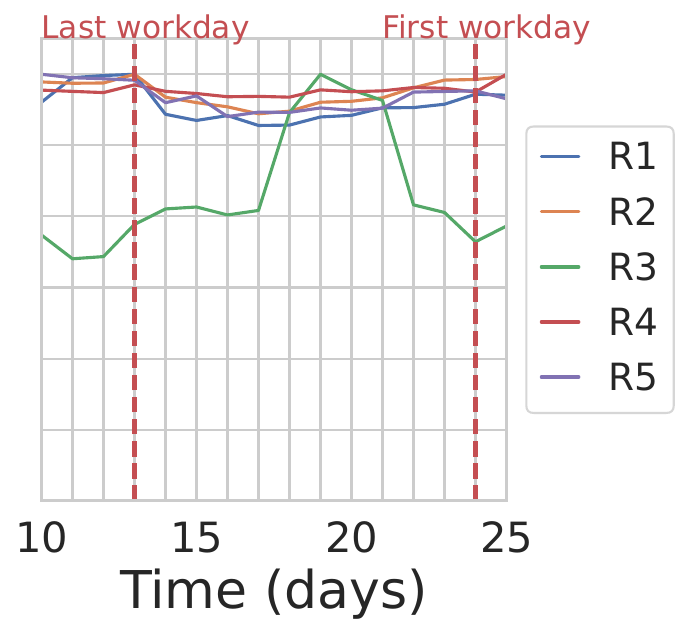}
                \label{fig:percent_change_cpu_usage_avg}}
            \caption{Normalized number of pods and allocated CPU during holiday period. Day 13 is the last working day before the holiday. Day 24 is the first working day after the holiday.}
            \label{fig:percent_change_holiday}
        \end{center}   
\end{figure}

Figure \ref{fig:peak_trough_ratio_vs_cold_starts_all_regions} shows the peak-to-trough ratio of each function against the total number of cold starts for that function over the duration of our dataset. We observe a concentration of functions with larger peak-to-trough ratios and a higher number of cold starts. Additionally, we see a cluster of functions with peak-to-trough ratios close (or equal) to 1, that tend to have fewer than 44640 cold starts (equivalent to one cold start per minute for 31 days), which corresponds to the same group of infrequently invoked functions described in Figure \ref{fig:peak_trough_ratio_vs_sum_median_avg_day_all_regions}. Hence, we see that high numbers of cold starts are a result of workloads with large fluctuations in their invocation patterns leading to frequent autoscaling decisions, or a large number of functions that are all invoked at most once per minute, falling just outside of the pod keep alive time. A successful policy to minimize the number of cold starts will address both of these sources.

\paragraph{Holidays}
Holidays and other special occasions may have a significant impact on cloud traffic. Certain workloads may increase or decrease during and around holiday periods~\cite{huang2022metastable}, while other workload types (e.g. associated with shopping and commercial loads) increase before holidays~\cite{bodik2010characterizing}. Our dataset includes a week-long holiday period. To the best of our knowledge, this is the first serverless dataset that includes such patterns.

Figure \ref{fig:percent_change_holiday} shows the number of pods and mean CPU usage, normalized to their maximum value during the same number of days before the holiday. The last working day before the holiday is on day 13, and the first working day after the holiday is day 24.
Both Figures \ref{fig:percent_change_num_pods} and \ref{fig:percent_change_cpu_usage_avg} show a similar pattern. Regions 1, 2, 4, and 5 all peak on day 13, decrease during the holiday, and increase to another peak on day 24. This suggests a pre-holiday `rush' and post-holiday `catch-up'. Region 3 experiences a different pattern, where its workload increases substantially at the start of the holiday, and reduces again towards the end.

\begin{tcolorbox}[title={\centering Complex origin of cold starts},colframe=customgrey, coltitle=black]
High numbers of cold starts can be contributed both by workloads with large fluctuations in invocation patterns, or simply by a large number of functions that are invoked with a periodicity greater than one minute, falling just outside the pod keep alive time. Some functions have peak-to-trough ratios greater than 1000. A successful policy to minimize the number of cold starts will address both of these sources. Holidays usually decrease resource allocation, with some regional variations.
\end{tcolorbox}

\begin{figure*}[h]
     \begin{center}

            \subfloat[Running pods per hour by trigger type.]{
                \includegraphics[height=5.4cm]{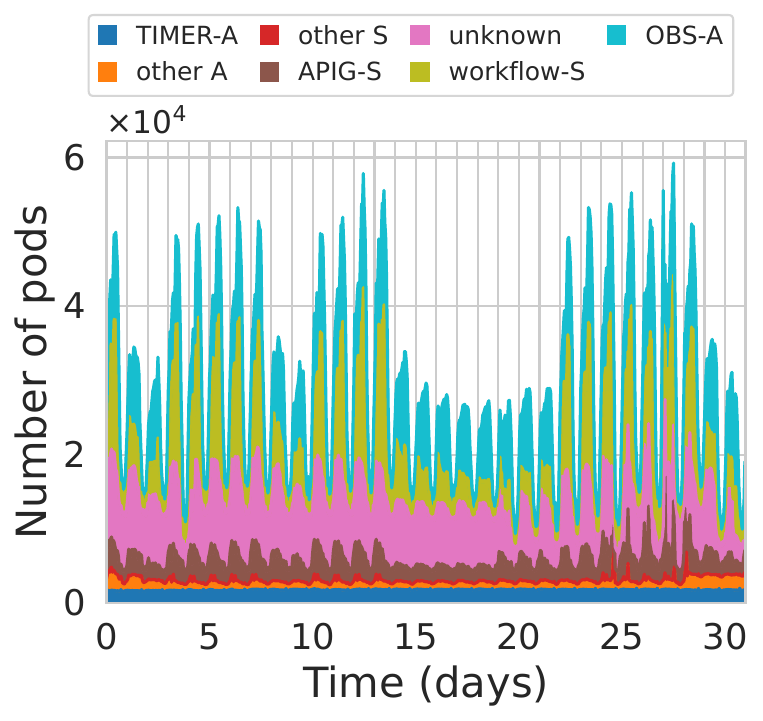}\label{fig:triggerTypeTS}}
            \hfill
            \subfloat[Running pods per hour by runtime.]{
                \includegraphics[height=5.4cm]{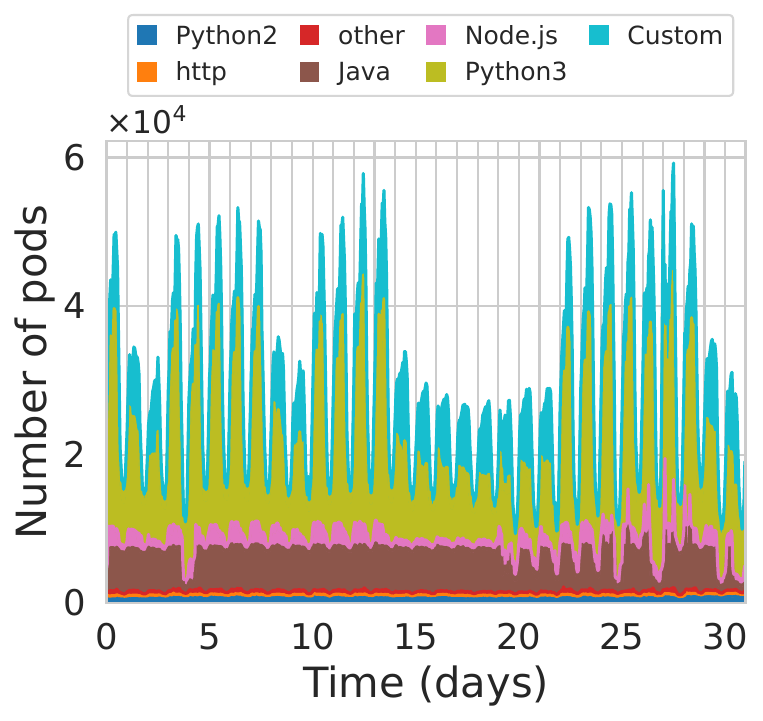}
                \label{fig:podsTS}}
            \hfill
            \subfloat[Running pods per hour by resource allocation.]{
                \includegraphics[height=5.4cm]{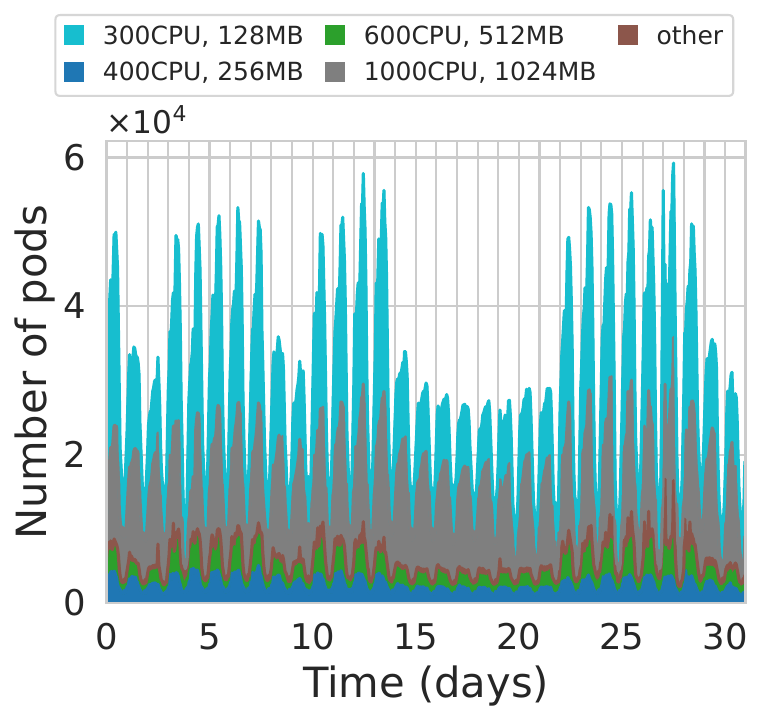}
                \label{fig:timeseries_stacked_grouped_by_cpu_mem_R2}}
            \hfill
            \subfloat[Proportions grouped by trigger type.]{
                \includegraphics[height=4.3cm]{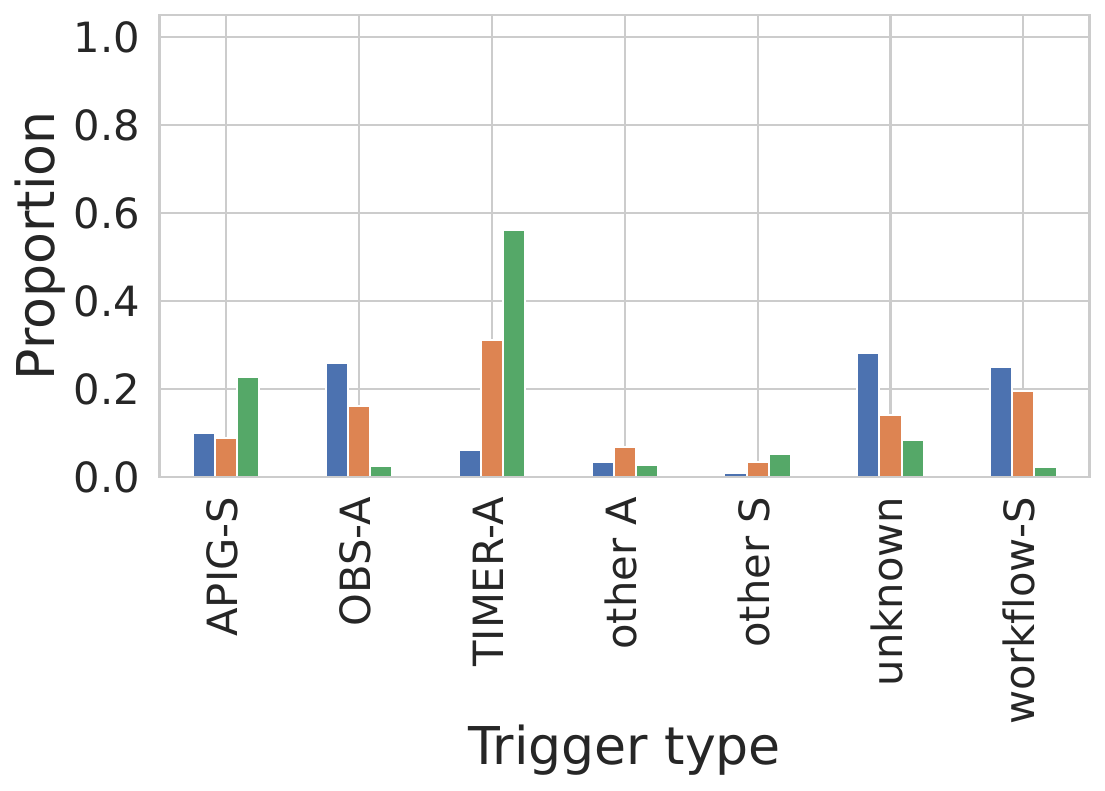}
                \label{fig:bar_proportions_grouped_by_triggerType_invocationType}}
            \hspace{3mm}
            \subfloat[Proportions grouped by runtime.]{
                \includegraphics[height=4.3cm]{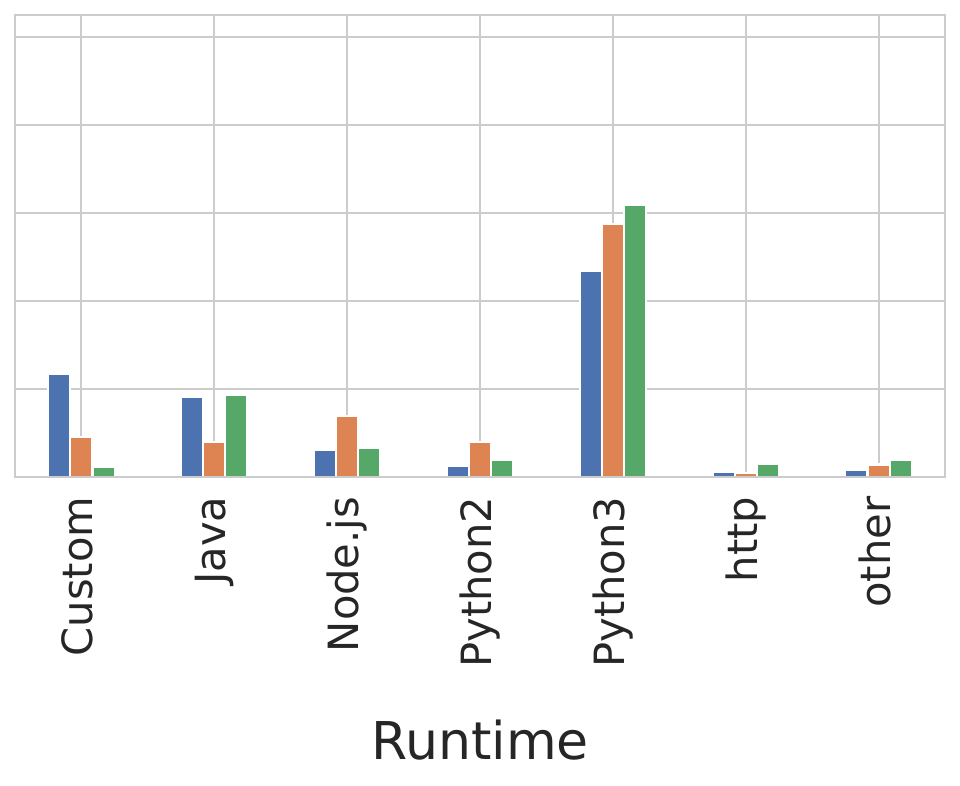}
                \label{fig:bar_proportions_grouped_by_runtime}}
            \hfill
                \subfloat[Proportions grouped by resource allocation.]{
                \includegraphics[height=4.3cm]{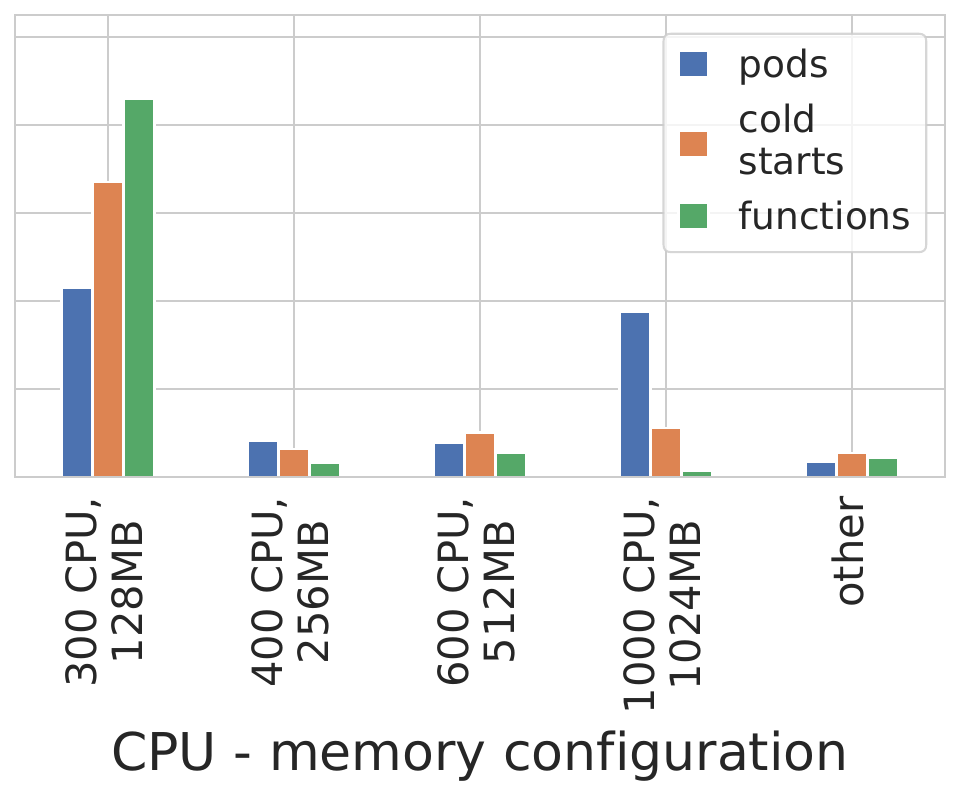}
                \label{fig:PodTriggerStats}}
            \caption{Proportions of running pods, cold starts, and functions by trigger, runtime, and resource allocation in Region 2.}
            \label{fig:bar_proportions}
        \end{center}
\end{figure*}

\begin{figure}[h]
	\centering
        \includegraphics[width=0.6\linewidth]{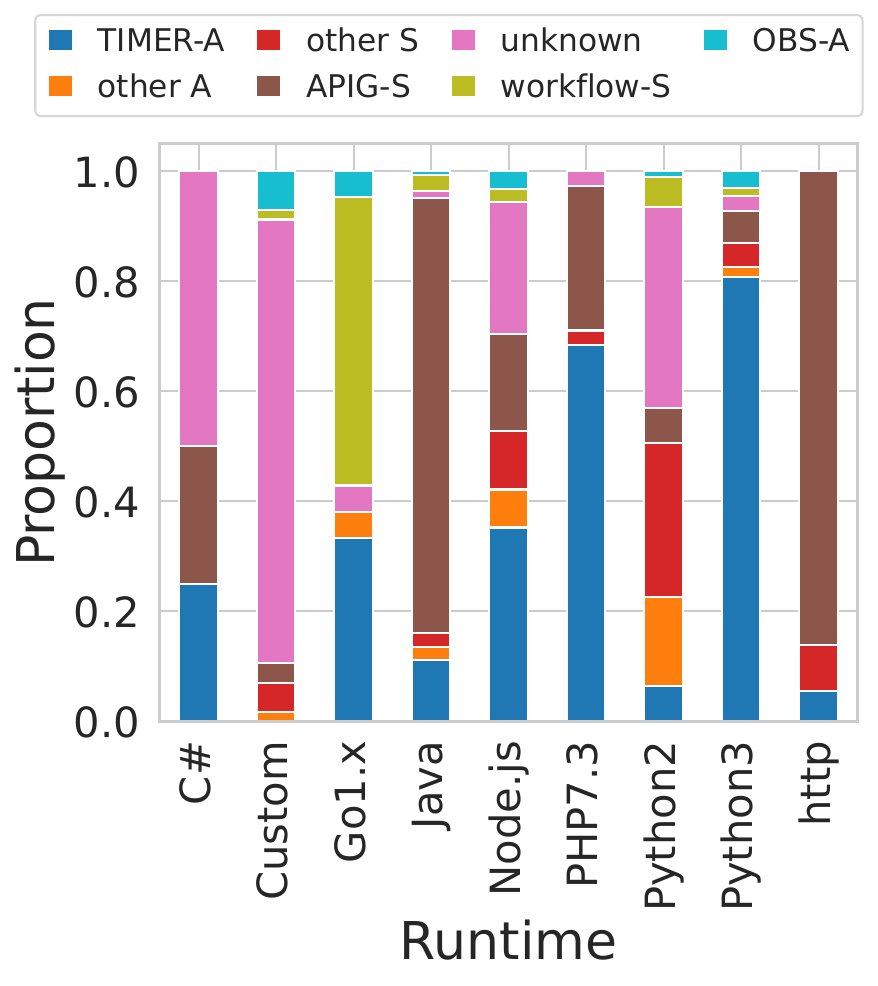}  
	\caption{Region 2 trigger types by runtime.}
    \label{fig:bar_stacked_runtimes_triggerTypes}
\end{figure}

\subsection{Runtimes and trigger types}\label{section:runtimes_and_trigger_types}

We now describe the distributions of trigger types, runtimes, and their combinations focusing on Region 2. A function can use one of preinstalled runtimes or use a custom image. Preinstalled runtimes include $C\#$, $Go1.x$, $Java$, $NodeJS$, $PHP7.3$, $Python2$ (for legacy reasons), $Python3$, and $HTTP$. When deployed, a function is invoked using one of many triggers. Some trigger types can only be invoked asynchronously, while others can be invoked synchronously. For synchronous (S) requests, the invoking program waits for a response, while for asynchronous (A) requests the invoking program does not wait for a response, typically checking the result later.
Our platform supports the following trigger types: 
\begin{enumerate}
    \item \textbf{API gateway (APIG) (S and A)} is an API hosting service. The function can be invoked through HTTPS using a custom REST API and a specified backend;
    \item \textbf{Timer} invokes functions based on a cron-style timer;
    \item \textbf{Cloud Trace Service (CTS) (A only)} is a platform monitoring service;
    \item \textbf{Data Ingestion Service (DIS) (A only)} triggers functions by a data stream, e.g. to process updated records; 
    \item \textbf{Log Tank Service (LTS) (A only)} triggers functions by logging events; 
    \item \textbf{Object Storage Service (OBS) (A only)} invokes functions by storage events, e.g. object creation or deletion; 
    \item \textbf{Simple Message Notification (SMN) (A only)} triggers functions using messages posted under topics; 
    \item \textbf{Kafka} triggers functions using a Kafka queue; 
    \item \textbf{Workflow (S and A)} lets functions directly trigger other functions.
\end{enumerate}

In addition, our system supports a function to have multiple trigger types. Since some of these trigger types are seldom used, we aggregate them in our analysis. We aggregate all trigger types except for timers, OBS-A, APIG-S, and workflow-S. We split the aggregation into other synchronous (other S) or other asynchronous (other A). Timers account for 42\% of all triggers, followed by APIG-S (23\%), APIG-S and TIMER-A combination (13\%), with the remaining triggers and trigger type combinations representing less than 5\% of functions each. The majority of functions only have one trigger type, with only a handful having two or more trigger types.

\paragraph{Trigger types by runtime.} To analyze the relationship between runtime and trigger type, Figure~\ref{fig:bar_stacked_runtimes_triggerTypes} shows a stacked bar chart of trigger types as proportions of the total number of functions for a given runtime. We note that the prevalence of different trigger types varies considerably between runtimes. For example, $Python3$, $PHP7.3$, and $Node.js$ functions are mostly triggered by timers, while $Java$ and $HTTP$ runtimes tend to use APIG-S triggers. Asynchronous triggers other than OBS and timers are most significant in $Python2$, and even then only account for a minority of functions. We note that a small proportion of our data does not have the runtime or the trigger types logged.

\begin{figure*}[h]
        \begin{center}
            \subfloat[Cold start times per region.]{
                \includegraphics[height=3.5cm]{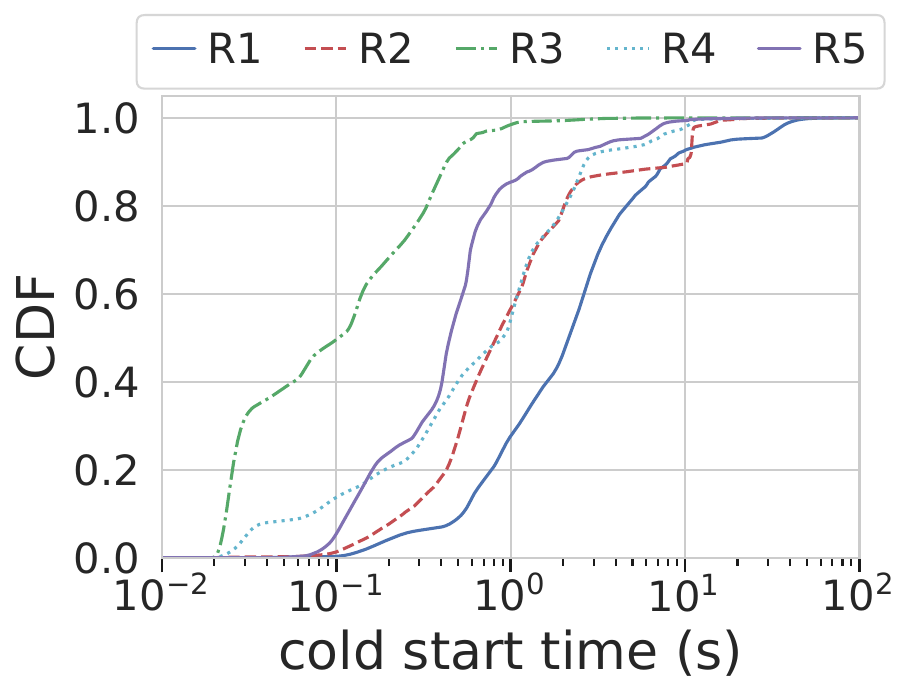}
                \label{fig:cdf_cold_starts_per_region}}
            \hfill
            \subfloat[Fit for cold start times.]{
                \includegraphics[height=3.5cm]{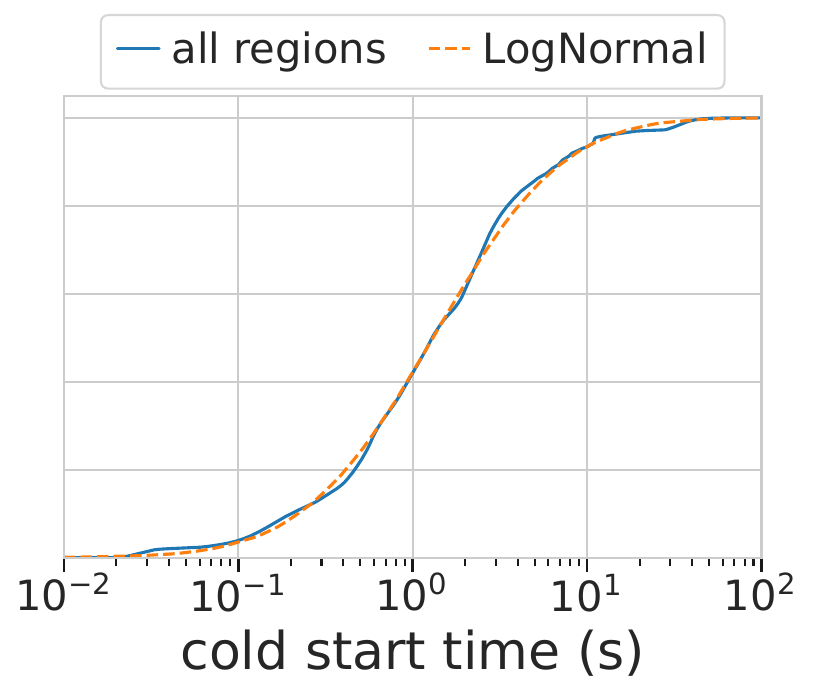}
                \label{fig:cold_starts_cdf_fitted}}
            \hfill
            \subfloat[CDF of inter-arrival times.]{
                \includegraphics[height=3.5cm]{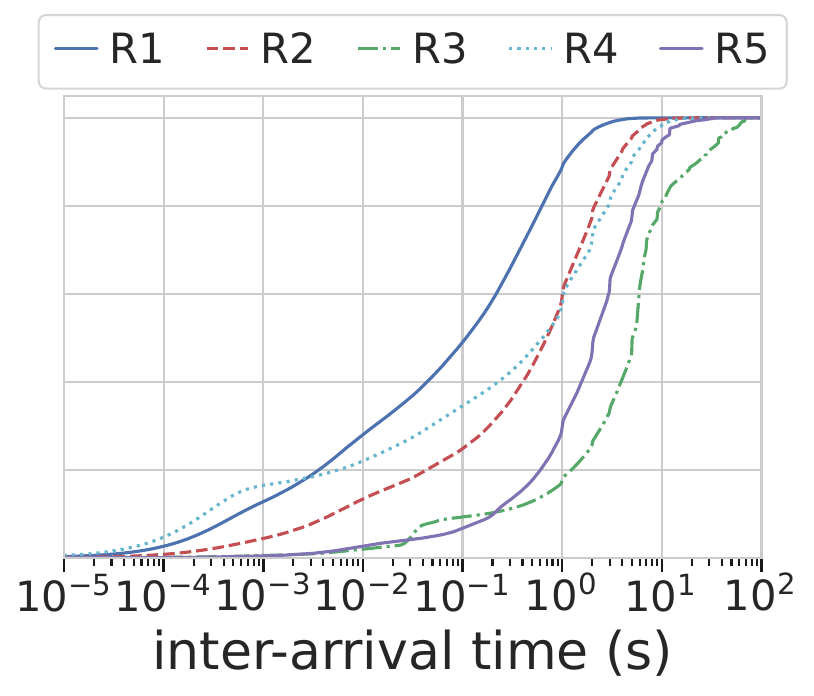}
                \label{fig:cdf_iat}}
            \hfill
            \subfloat[Fit for inter-arrival times.]{
              \includegraphics[height=3.5cm]{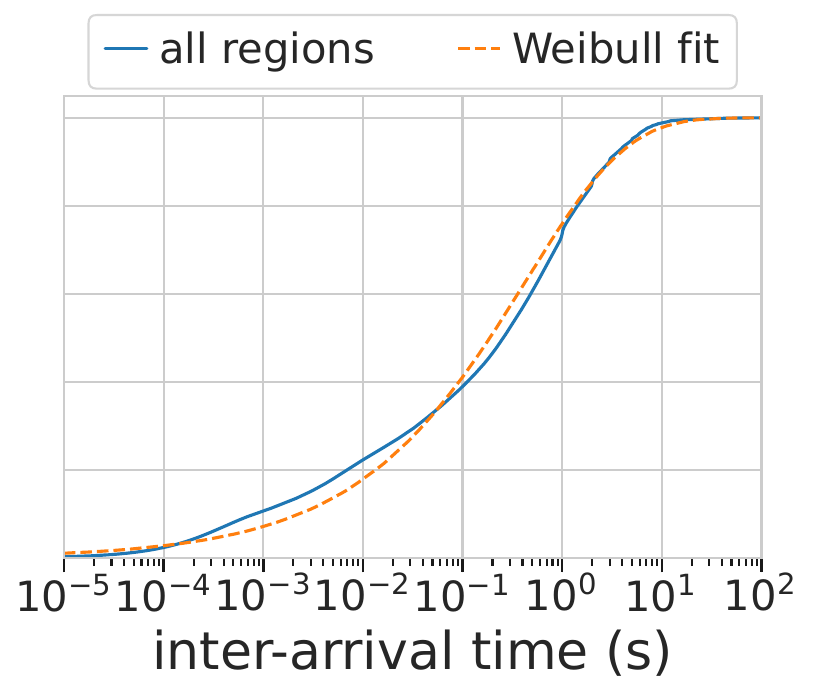}
                \label{fig:iat_cdf_fitted}}
            \caption{CDFs and distribution fits for cold start times and inter-arrival times between cold starts. LogNormal fit for cold starts has mean 3.24 and standard deviation 7.10. Weibull fit for inter-arrival times has mean 1.25 and standard deviation 3.66.}
            \label{fig:cold_start_cdfs_and_fits}
        \end{center}   
\end{figure*}

\paragraph{Pods by runtime and trigger type over time.} 

Since invocations have periodic patterns, it is worth investigating if a similar periodicity can be observed in the number of running pods when grouped by runtime language or trigger type. Such patterns may be useful in workload collocation, pre-warming, and scheduling.
Figures~\ref{fig:triggerTypeTS} and ~\ref{fig:podsTS} shows the average number of pods per hour in Region 2 by trigger type and runtime for the full duration of our dataset. While some runtimes and trigger types have a stable number of pods, others show significant variability. For example, $Python3$ runtimes and workflow-S trigger types show large variability that correlates with time of day. Both peak during working hours, and decrease during nighttime. Notably, Figure~\ref{fig:triggerTypeTS} shows that the number of pods allocated for timers does not vary much with time, even though they represent almost 60\% of functions as shown in Figure~\ref{fig:bar_proportions_grouped_by_triggerType_invocationType}. Many timer functions tend to be invoked infrequently (at most once per minute) and are only allocated a single pod. 

Daily periodicities in public cloud workloads have previously been attributed to timers and diurnal user patterns ~\cite{Joosen_2023}. Our analysis suggests that timer workloads do not contribute significantly to the total platform load periodicity, but that it is user-driven diurnal behaviour driven by APIG calls or workflows that contribute more to daily periodicity.
Such predictable patterns may be exploited by serverless management systems and cluster managers to reduce the number and duration of cold starts by prewarming instances of these runtimes according to these patterns.
We also note that most trigger types and runtimes have strong weekly periodicity, with approximately 30\% more pods allocated during weekdays compared to weekends. This weekly periodicity is interrupted during the holiday period beginning on day 14, with day 13 being the last working day, with workload level similar to weekend levels. Timer-triggered functions are almost completely unaffected by the holiday period. 

Synchronous trigger types, such as workflow-S and APIG-S, have very strong daily periodicity and contribute significantly to peaks in the overall number of allocated pods on the platform. Given that these requests are synchronous, there may not be significant potential for flexible scheduling. However, OBS triggered functions, which are asynchronous, also have very strong daily oscillations and contribute significantly to the peak time pod allocations, and account for almost 30\% of running pods, as seen in Figure~\ref{fig:bar_proportions_grouped_by_triggerType_invocationType}. Asynchronous triggers may be used for tasks that are not latency critical, such as log batch analysis triggered by a new logging event (LTS) or the presence of new files (OBS). Hence, if the provider can differentiate between asynchronous triggers that have less latency critical deadlines versus asynchronous requests that are time-critical but where the system does not wait for the response, peak-shaving can be used, whereby the allocation of these pods and execution of these requests is delayed. Given the narrow peak widths, even a short delay could significantly reduce peak pod allocations.

Finally, we note the changing periodicity characteristics of certain workloads. For example, the number of pods serving $Java$ functions varied very little until day 18, after which a strong diurnal periodicity began. This means that a serverless platform must be able to detect these changes in the workload characteristics as some of these changes, if well detected, can present new opportunities for online system optimizations.

\paragraph{Pods by CPU and memory allocation.}

Figure \ref{fig:timeseries_stacked_grouped_by_cpu_mem_R2} shows the number of running pods over time grouped by their resource allocation. We can see that these different resource configurations contribute different amounts to the total periodic fluctuations. As described in Section \ref{section:cold_starts_background}, our platform maintains pools of inactive pods to be used as demanded by user traffic. The predictability of these patterns may allow the provider to maintain just enough pods to meet expected demand, while avoiding excessive overallocation using online predictors and dynamic resource pre-warming.

\begin{tcolorbox}[title={\centering Predictive scheduling and peak shaving},colframe=customgrey, coltitle=black]
Function invocations follow periodic patterns that could be leveraged to pre-warm pods with popular configurations, thus reducing cold starts. Delaying pod allocation for asynchronously invoked functions could reduce peaks if they are not latency critical.
\end{tcolorbox}

\section{Causes of cold starts}\label{section:where_can_we_find_cold_starts}

We now conduct an in-depth analysis of cold starts. We first study cold starts and their components in all regions. We then study Region 2 in further depth, examining how cold starts vary by trigger type, runtime, and resource allocation. 

\subsection{Cold start distributions} 
We start our analysis by plotting CDFs of the duration and inter-arrival times of cold starts. Figure \ref{fig:cdf_cold_starts_per_region} shows the distribution of cold start times for different regions. We see large variations in cold start times between different regions, with medians between 0.1 seconds and 2 seconds. We see that cold start times in all regions have a long tail. Figure \ref{fig:cdf_iat} shows the distribution of inter-arrival times (IAT) between cold starts for different regions. Median inter-arrival times range from 0.1 seconds in R1 to several seconds in R3. 

To capture the average behavior across regions, e.g. for simulation purposes, we fit a distribution to all cold start times and another distribution for cold start inter-arrival times. Cold start times can be approximated with a LogNormal distribution, while inter-arrival times can be approximated with a Weibull distribution, a common distribution for modelling event inter-arrival times~\cite{weibull}. Figure \ref{fig:cold_starts_cdf_fitted} shows the distribution of cold start times across all regions with a LogNormal fit with mean 3.24 and standard deviation 7.10, and Figure \ref{fig:iat_cdf_fitted} shows inter-arrival times for all regions with a Weibull fit with mean 1.25 and standard deviation 3.66. These fits are very close to the measured data from our system. We believe that researchers working on problems related to cold start time optimizations can use these fits to run simulation experiments for cold start optimizations.

\subsection{Components of cold start times}

\begin{figure*}
        \begin{center}
            \subfloat[Region 1]{
                \includegraphics[height=3.62cm]{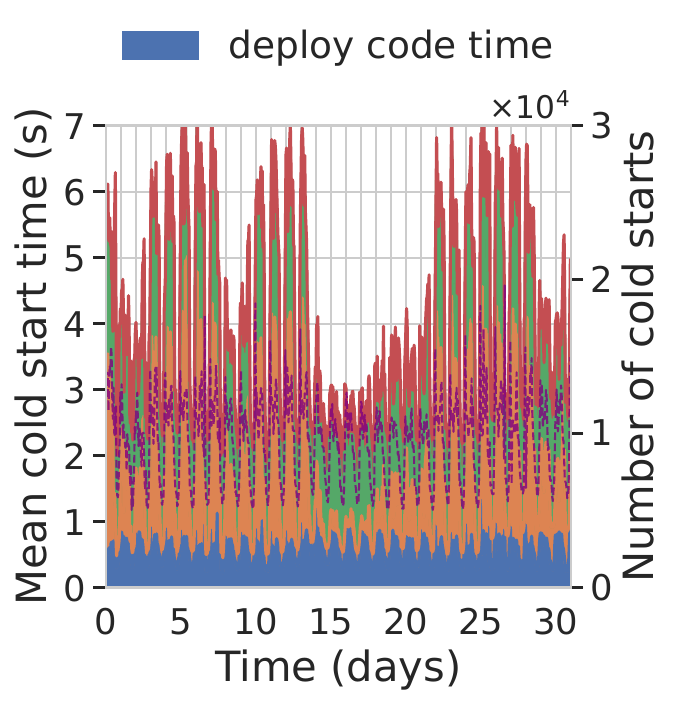}
                \label{fig:timeseries_cold_start_stacked_R1}}
                \hspace{-2mm}
            \subfloat[Region 2]{
              \includegraphics[height=3.62cm]{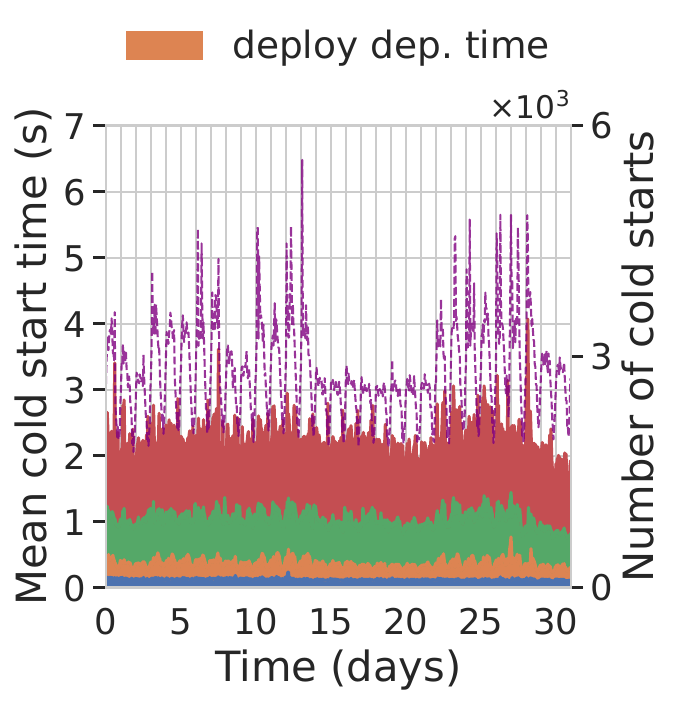}
                \label{fig:timeseries_cold_start_stacked_R2}}
                \hspace{-2mm}
            \subfloat[Region 3]{
              \includegraphics[height=3.62cm]{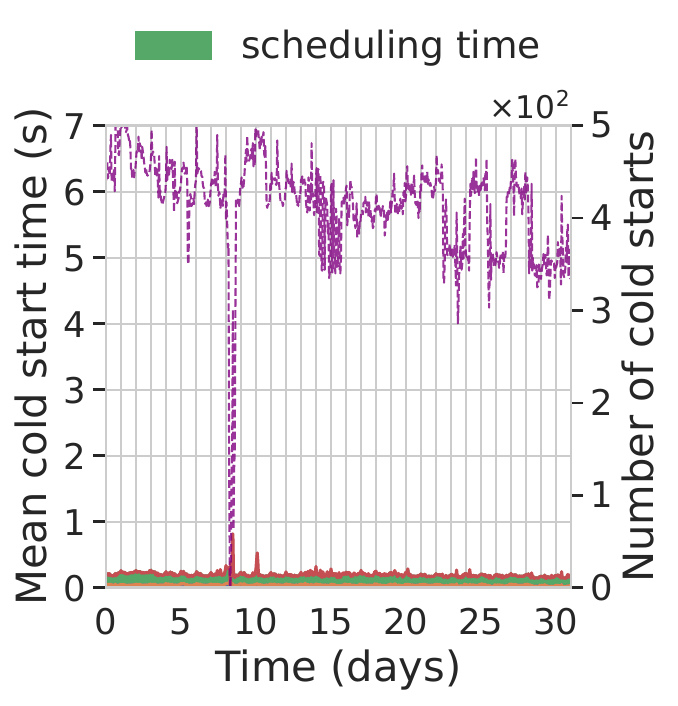}
                \label{fig:timeseries_cold_start_stacked_R3}}
                \hspace{-2mm}
            \subfloat[Region 4]{
              \includegraphics[height=3.62cm]{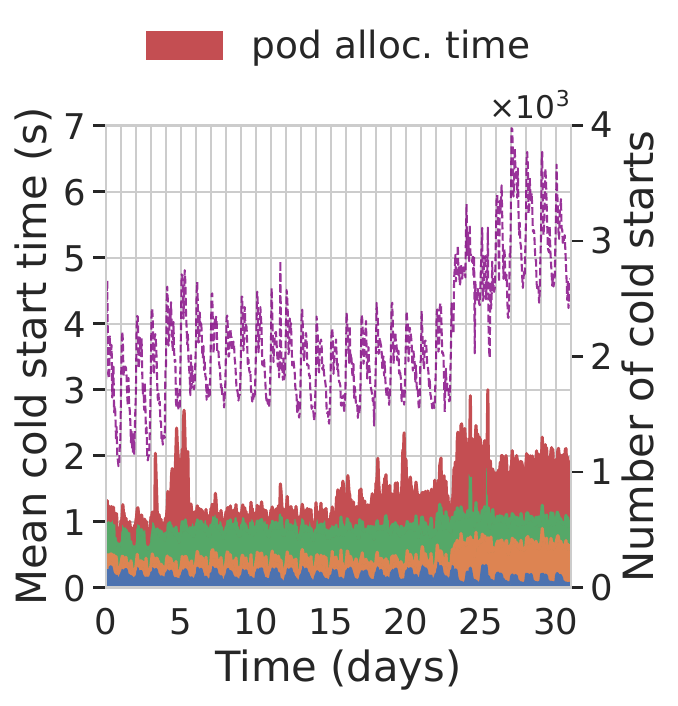}
                \label{fig:timeseries_cold_start_stacked_R4}}            
                \hspace{-2mm}
            \subfloat[Region 5]{
              \includegraphics[height=3.62cm]{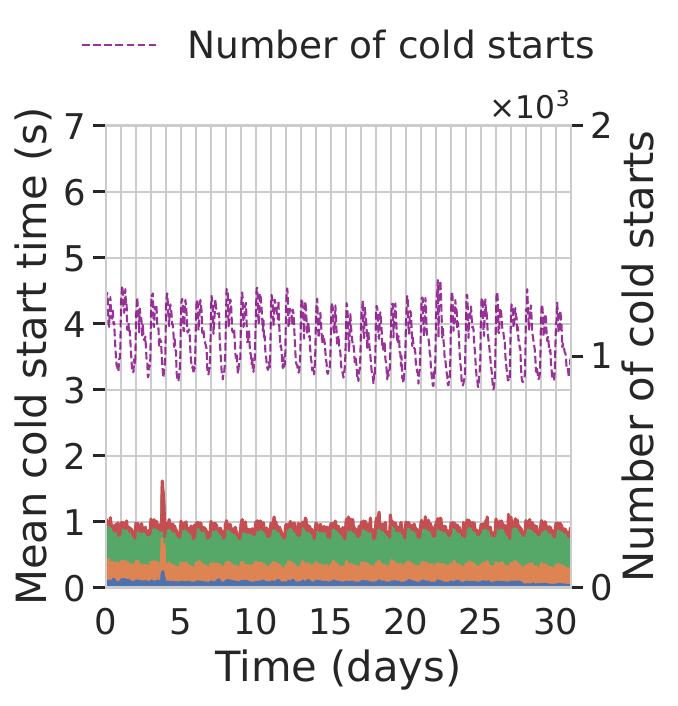}
                \label{fig:timeseries_cold_start_stacked_R5}}               
            \caption{Mean cold start time per hour split by components on left axis. Number of cold starts per hour on right axis.}
            \label{fig:stacked_bar_cold_start_time_components} 
        \end{center}
\end{figure*}

\begin{figure*}
        \begin{center}
            \subfloat[Region 1]{
                \includegraphics[height=3.8cm]{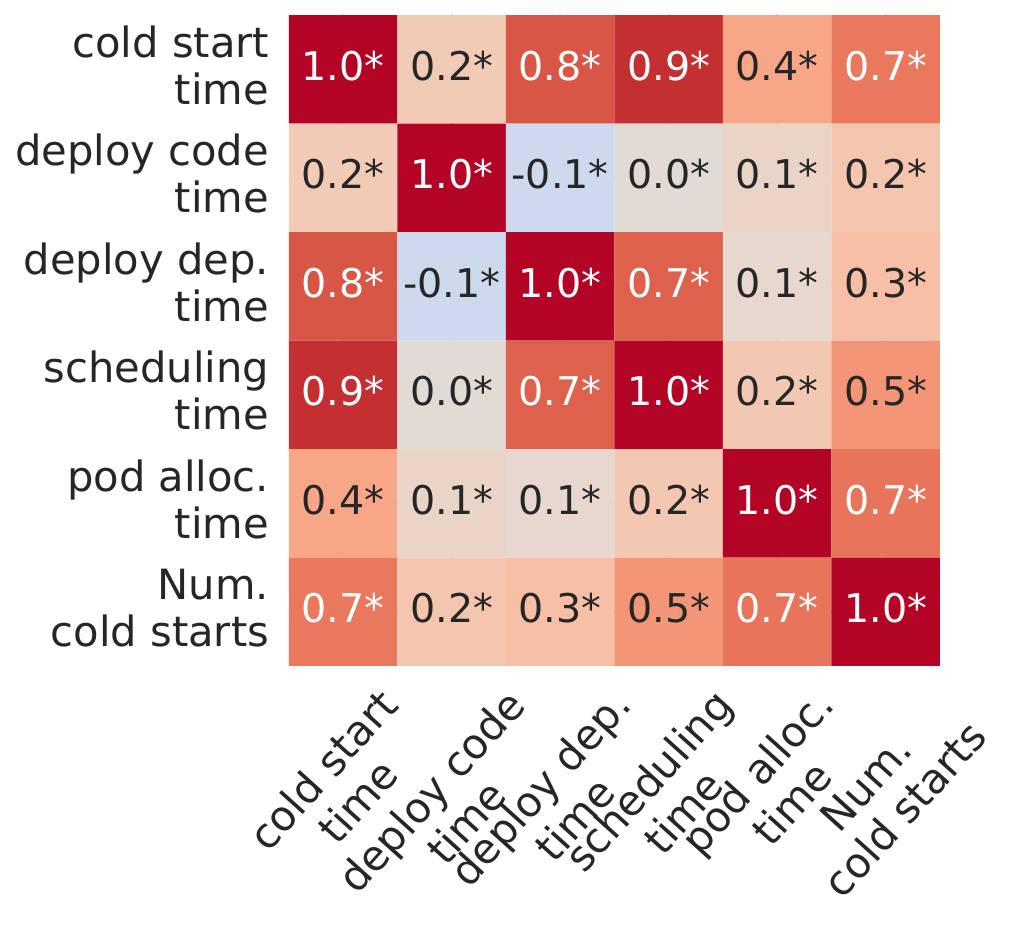}
                \label{fig:correlation_cold_start_R1}}
                \hspace{-2mm}
            \subfloat[Region 2]{
              \includegraphics[height=3.8cm]{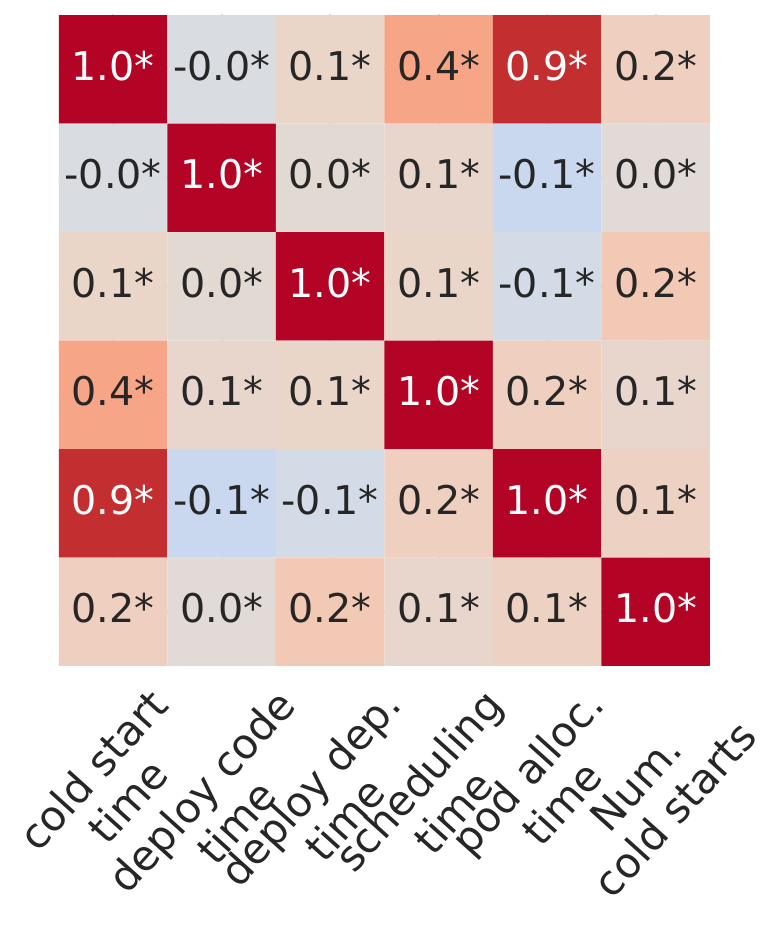}
                \label{fig:correlation_cold_start_R2}}
                \hspace{-2mm}
            \subfloat[Region 3]{
              \includegraphics[height=3.8cm]{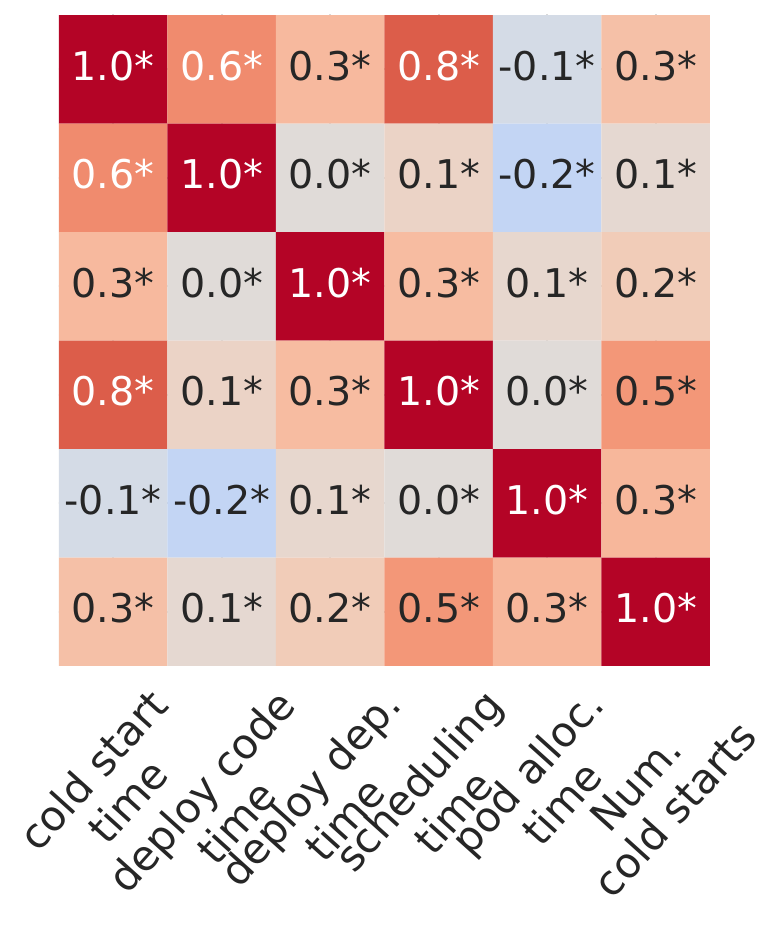}
                \label{fig:correlation_cold_start_R3}}
                \hspace{-2mm}
            \subfloat[Region 4]{
              \includegraphics[height=3.8cm]{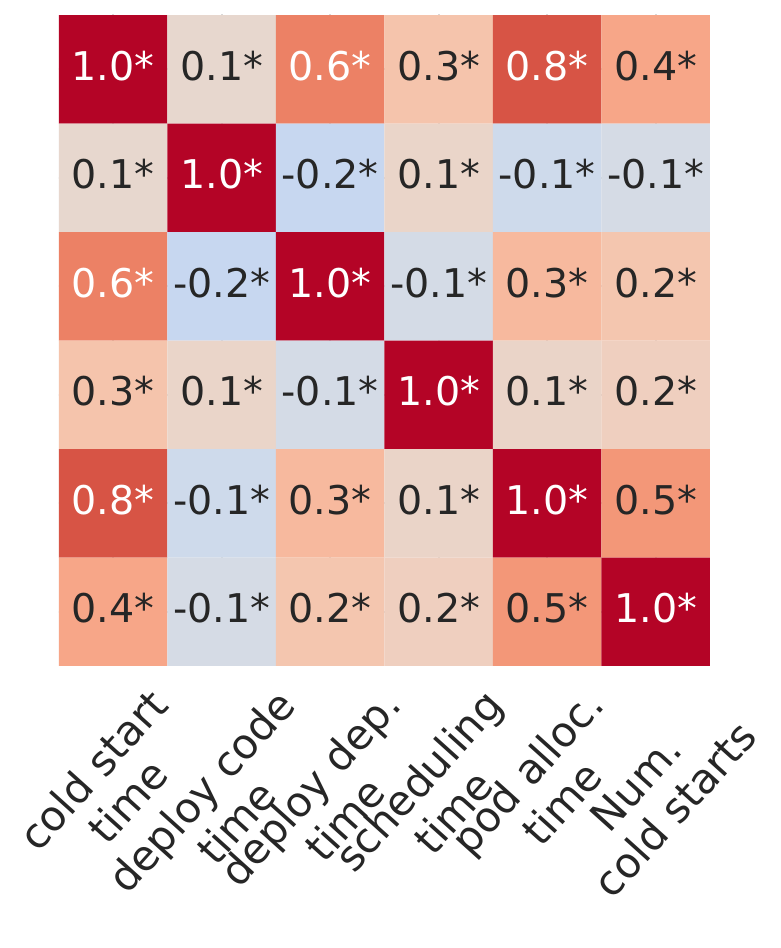}
                \label{fig:correlation_cold_start_R4}}            
                \hspace{-2mm}
            \subfloat[Region 5]{
              \includegraphics[height=3.8cm]{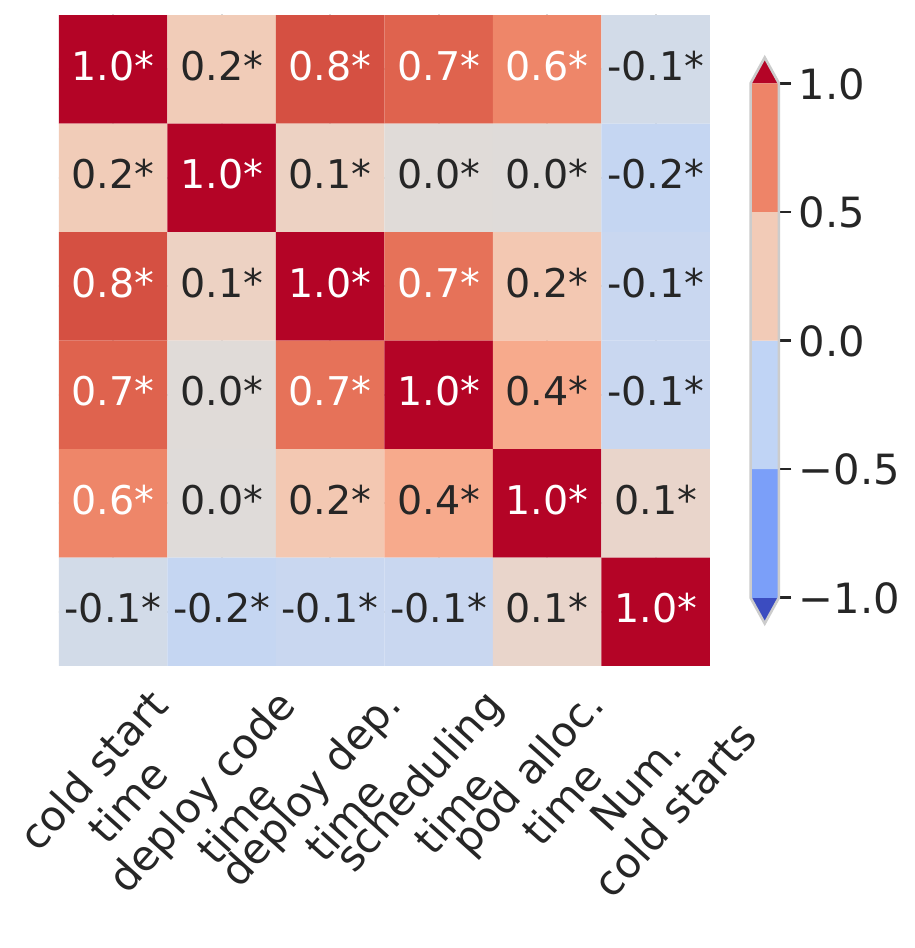}
                \label{fig:correlation_cold_start_R5}}               
            \caption{Spearman correlations of mean cold start time components per minute aggregated over all functions for each region. Correlation values where $p<0.05$ are marked with an asterisk.}
            \label{fig:cold_start_correlations} 
        \end{center}
\end{figure*}

When a cold start occurs, several steps are required until the new pod is operational. We log the time taken for each step taken during a cold start, namely; the time taken to start a pod if no free pods exist or to select a pod from the existing pool to be used by the newly started function (referred to as \textit{pod allocation time}); the time taken to download, extract, and deploy function code (referred to as \textit{deploy code time}); the time to fetch and load dependencies (referred to as \textit{deploy dependency time}); and the time for networking, routing, and scheduling overheads (referred to as \textit{scheduling time}).

Figure \ref{fig:stacked_bar_cold_start_time_components} shows the mean cold start time per hour for each region, with stacked areas representing component times along with the total number of cold starts per hour. We plot the time taken to allocate a pod, deploy code, deploy dependencies, and scheduling, which together add up to the total mean cold start time on the left-hand axis. We see that the relative proportion taken up by each component varies over time as well as between regions.

In absolute and relative terms, pod allocation time in Region 2 is much higher than in Region 1. Mean total cold start times vary between 3 seconds in Region 1 to less than 0.3 seconds in Region 3. Periodicity of components varies. For example, pod allocation time in Region 2 has the largest oscillations of all components in that region and is in phase with the number of cold starts. Time to deploy dependencies and scheduling time oscillate less. Meanwhile, time to deploy code remains almost constant. We can see a significantly different pattern in Region 1, where the time to deploy code and dependencies both oscillate significantly. 

Day 23 is the first working day after the holiday, and all regions show an increase in number and duration of cold starts then. This effect is especially pronounced in Regions 1, 2, and 4. These regions in particular also see a strong increase in time to deploy dependencies and pod allocation time. These may be caused by first-time function code deployments following a prolonged period of inactivity, as well as competition for a small number of reserved pods. 

The different composition of cold start times in different regions at different times of day may occur due to differing workloads between regions or due to architectural differences between data centers. Different data centers may use different architectures, network topologies, or other structural differences that cause bottlenecks in different parts of the system when scaling. Identifying which component to optimize must be done within the context of the data center's architecture as well as in-depth workload analysis.

\paragraph{Correlations between cold start time components.}

Figure \ref{fig:cold_start_correlations} shows Spearman correlation values between cold start time and its components, as well as the number of cold starts, averaged over all functions in that region. We can observe several trends that generalize across regions. Cold start time tends to be positively correlated with the number of cold starts, although the strength of this correlation varies across regions. Scheduling time and pod allocation time are positively correlated with the number of cold starts, especially in Regions 1, 3, and 4, suggesting that increasing demand for cold starts may cause delays in scheduling and pod allocation. There are also several more isolated, stronger correlations, such as those between scheduling and deploying dependencies in Regions 1 and 5, pointing to region-specific bottlenecks.

\begin{tcolorbox}[title={\centering Cold start components across regions and time},colframe=customgrey, coltitle=black]

Cold start times and the components that dominate them vary significantly between regions, pointing to workload differences and potential effects of hardware setups. Mean cold start time tends to correlate positively with number of cold starts. 

\end{tcolorbox}

\paragraph{Resource allocation and cold start time.}

\begin{figure*}[h]
    \begin{center}{
    \subfloat[Total cold start time.]{
       \includegraphics[height=3.16cm]{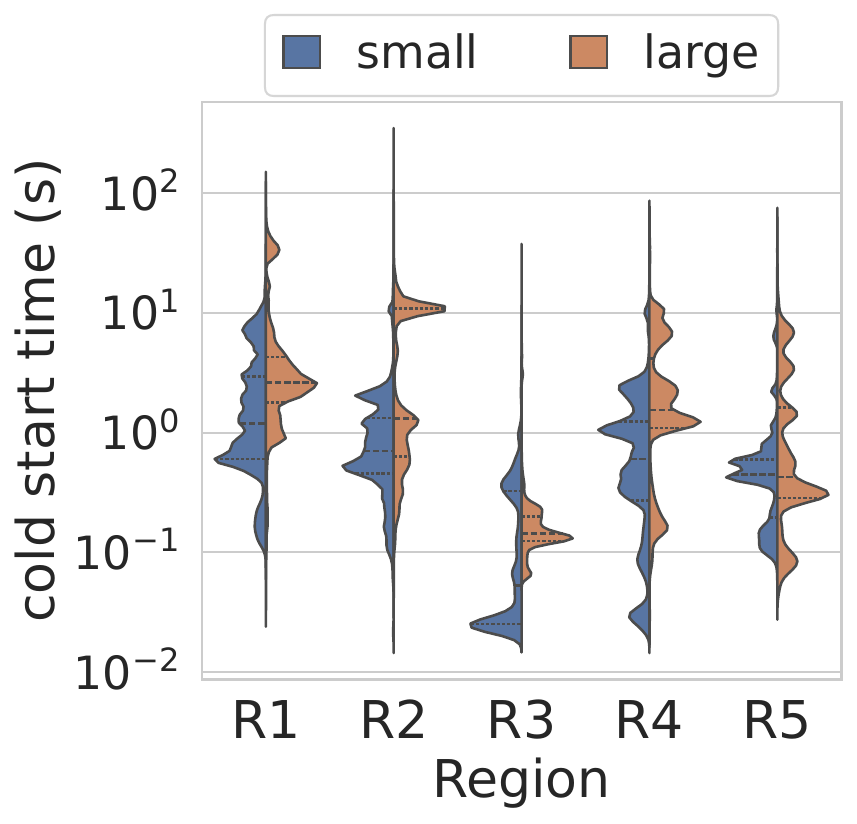}       \label{fig:cold_start_time_by_pool_violin_plot_totalCost_all_regions} 
    }
    \subfloat[Pod allocation time.]{ 
        \includegraphics[height=3.16cm]{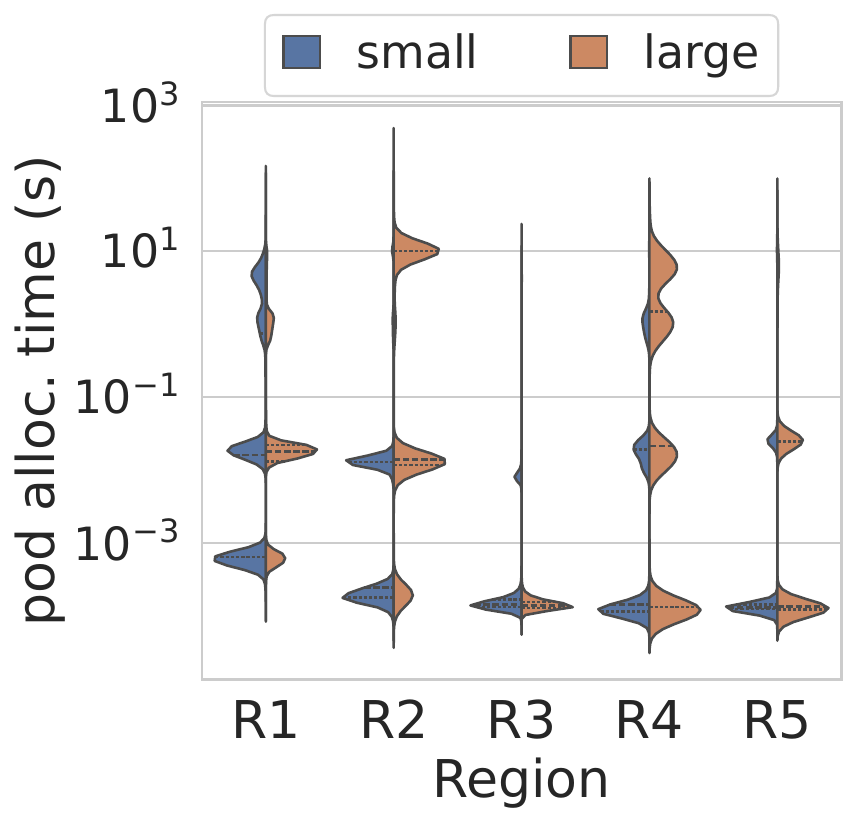}
        \label{fig:cold_start_time_by_pool_violin_plot_getpodfrompoolCost_all_regions} 
    }   
    \subfloat[Time to deploy code.]{
        \includegraphics[height=3.16cm]{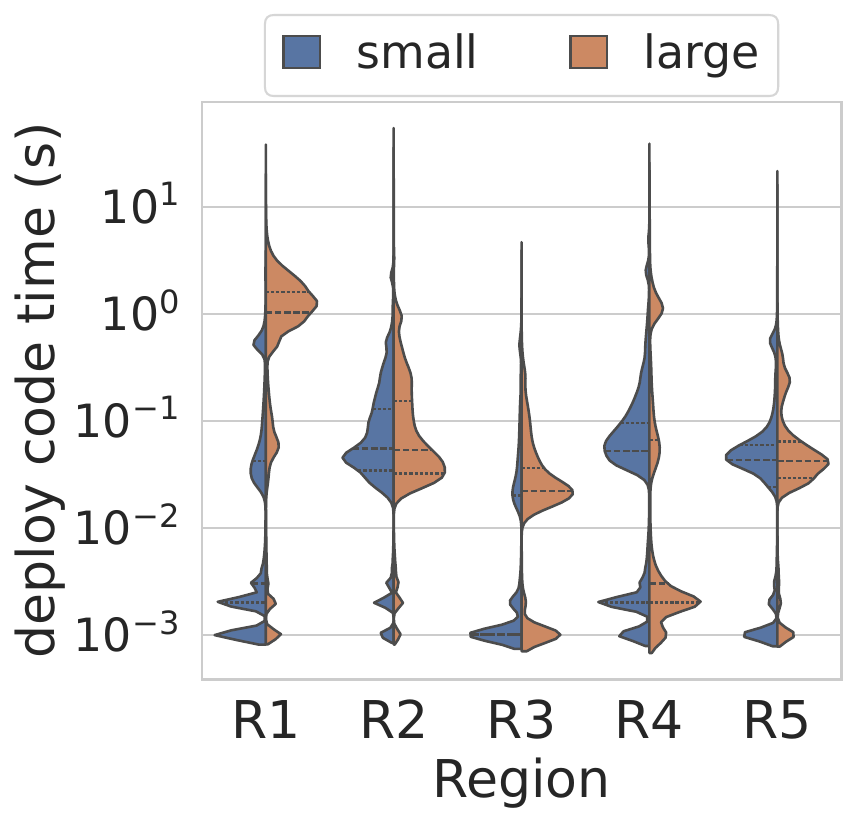}
        \label{fig:cold_start_time_by_pool_violin_plot_deployfunctioncompressedfileCost_all_regions} 
    }    
    \subfloat[Deploy dependency time.]{
       \includegraphics[height=3.16cm]{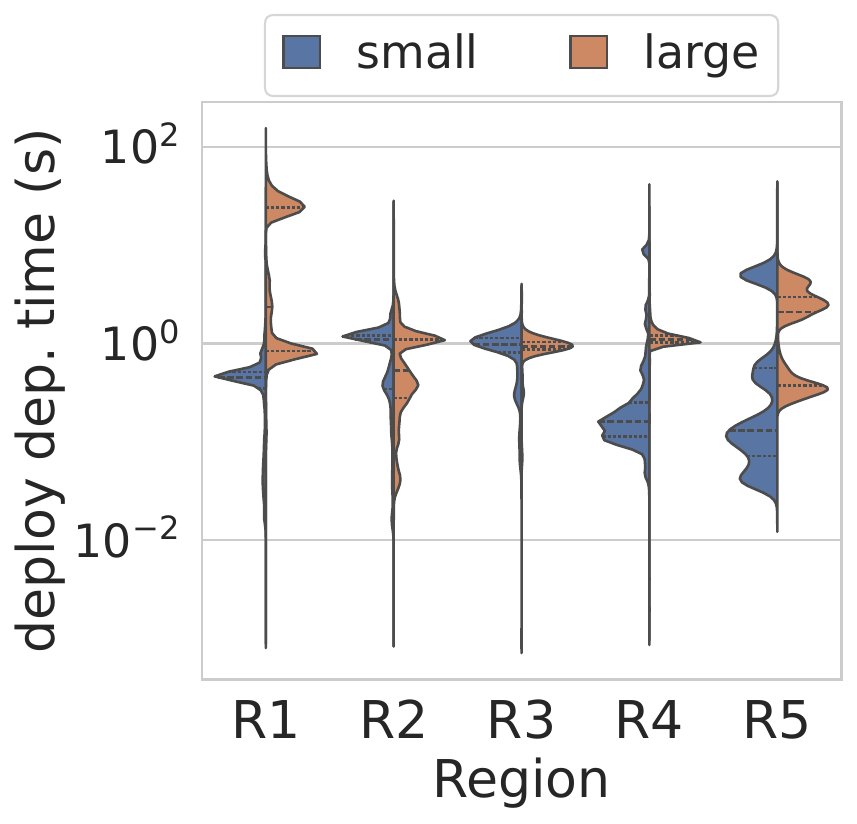}
       \label{fig:cold_start_time_by_pool_violin_plot_deploylayersCost_all_regions} 
    }
    \subfloat[Scheduling time.]{      
        \includegraphics[height=3.16cm]{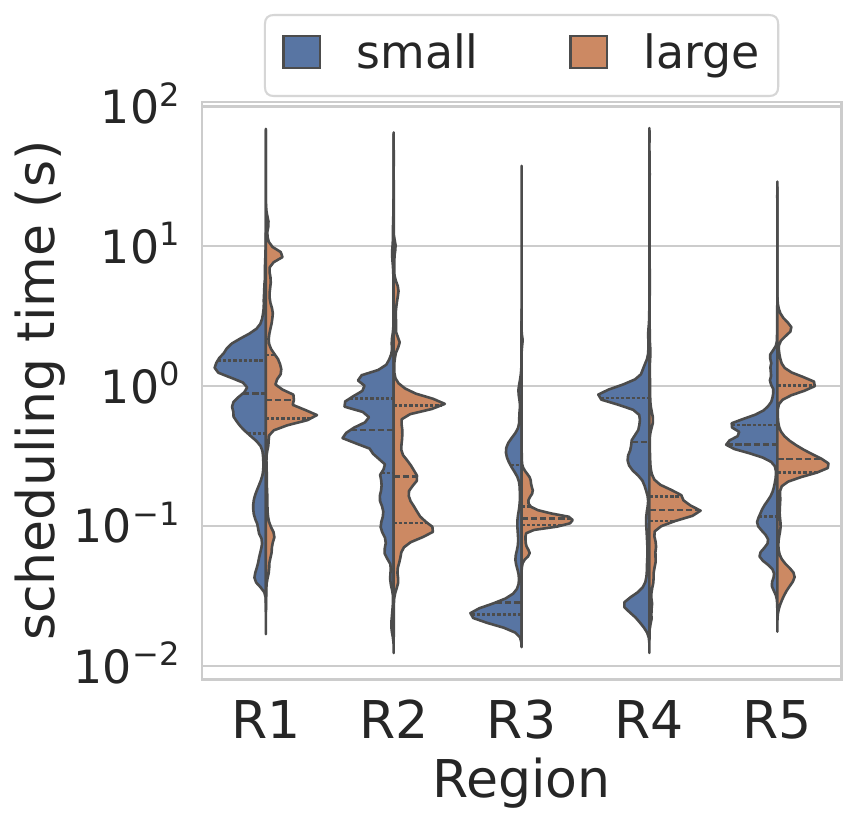}
        \label{fig:cold_start_time_by_pool_violin_plot_remainderCost_all_regions} 
    }
    }\end{center}
    \caption{Violin plot of cold start time and components by pool size for all regions with normalized width. For functions without layers, deploy dependency time is zero and excluded from plots. Dashed lines in the density plots represent quartiles.}
    \label{fig:cold_start_by_pool}  
\end{figure*}

Our system maintains pools of pods of different resource configurations, ranging from 300 millicores of CPU and 128MB of memory to 26 cores and 32GB of memory. We aggregate these pools into two groups: smaller pods with at most 400 millicores of CPU and 256MB of memory, and larger pods with allocations larger than that. In order to observe if there is a difference in the distribution of cold start times between small and large pods, Figure \ref{fig:cold_start_by_pool} shows a violin plot of the distributions of cold start times and their components for small and large resource pools. 
We observe that larger resource pools  tend to have longer median cold start times. The ratio of median cold start times between the high and low resource pods ranges from approximately 1:1 (Region 5) to 5:1 (Region 3). 

Figure~\ref{fig:cold_start_time_by_pool_violin_plot_getpodfrompoolCost_all_regions} shows the time to allocate a pod for large and small resource pools. During a cold start, the first step is to search for a pod of the appropriate configuration in a given resource pool. This search is conducted in stages, with the search expanding if a pod is not found. These stages clearly show themselves in the multimodal distribution in pod allocation time. For small resource pools, the pod allocation time tends to take less time, with a smaller proportion of cases expanding to further stages of the search with longer latency. For larger resource pools, the search tends to expand more frequently to later stages, taking a longer time to allocate a pod in total. This trend is consistent across all five regions.
Figure~\ref{fig:cold_start_time_by_pool_violin_plot_deployfunctioncompressedfileCost_all_regions} shows the time to deploy function code in small and large pods. We see that function code typically takes longer to deploy in larger pods, which is particularly pronounced in Regions 1, 2, 3, and 5. 
Figure~\ref{fig:cold_start_time_by_pool_violin_plot_deploylayersCost_all_regions} shows the time to deploy dependencies, if any. We see that time taken to deploy dependencies is longer for larger pods compared to smaller ones. Finally, Figure~\ref{fig:cold_start_time_by_pool_violin_plot_remainderCost_all_regions} shows the scheduling time in small and large pods. For regions 1, 3, and 4, smaller resource pods tend to have shorter scheduling times than larger pods. Regions 2 and 5 see the opposite pattern, where large pods have shorter scheduling times. Longer code and dependency deployment time may point to more complex code being deployed in larger pods.

\subsection{Which functions cause the most cold starts?}

We now focus our analysis on Region 2. We choose Region 2 as it has several interesting characteristics, such as large changes during the holiday period and large variations in different cold start components. Figures \ref{fig:bar_proportions_grouped_by_triggerType_invocationType}, ~\ref{fig:bar_proportions_grouped_by_runtime}, and~\ref{fig:PodTriggerStats} show the proportion of pods, cold starts, and number of functions accounted for by different trigger types, runtimes, and resource allocations respectively. The proportion of pods is calculated using the mean number of active pods per minute, while the proportion of cold starts is calculated using the number of newly started pods. Figure \ref{fig:bar_proportions_grouped_by_triggerType_invocationType} shows that different trigger types account for vastly different proportions of pod allocations, cold starts, and functions. For example, timers account for almost 60\% of functions and 30\% of cold starts, but only 5\% of running pods. Similarly, $Python3$ runtimes account for almost 50\% of all cold starts. For resource allocations, small CPU-memory allocations account for more than 60\% of cold starts. A provider can use the above data to decide the different percentages of pre-warmed runtimes and configurations. For example, since a large proportion of all functions are deployed as $Python3$ and also with a small CPU-Memory configuration, a provider can pre-warm a larger number of pods with $Python3$ deployed in small CPU-memory configuration pods. 



\begin{figure}
	\centering
    \includegraphics[width=1\linewidth]{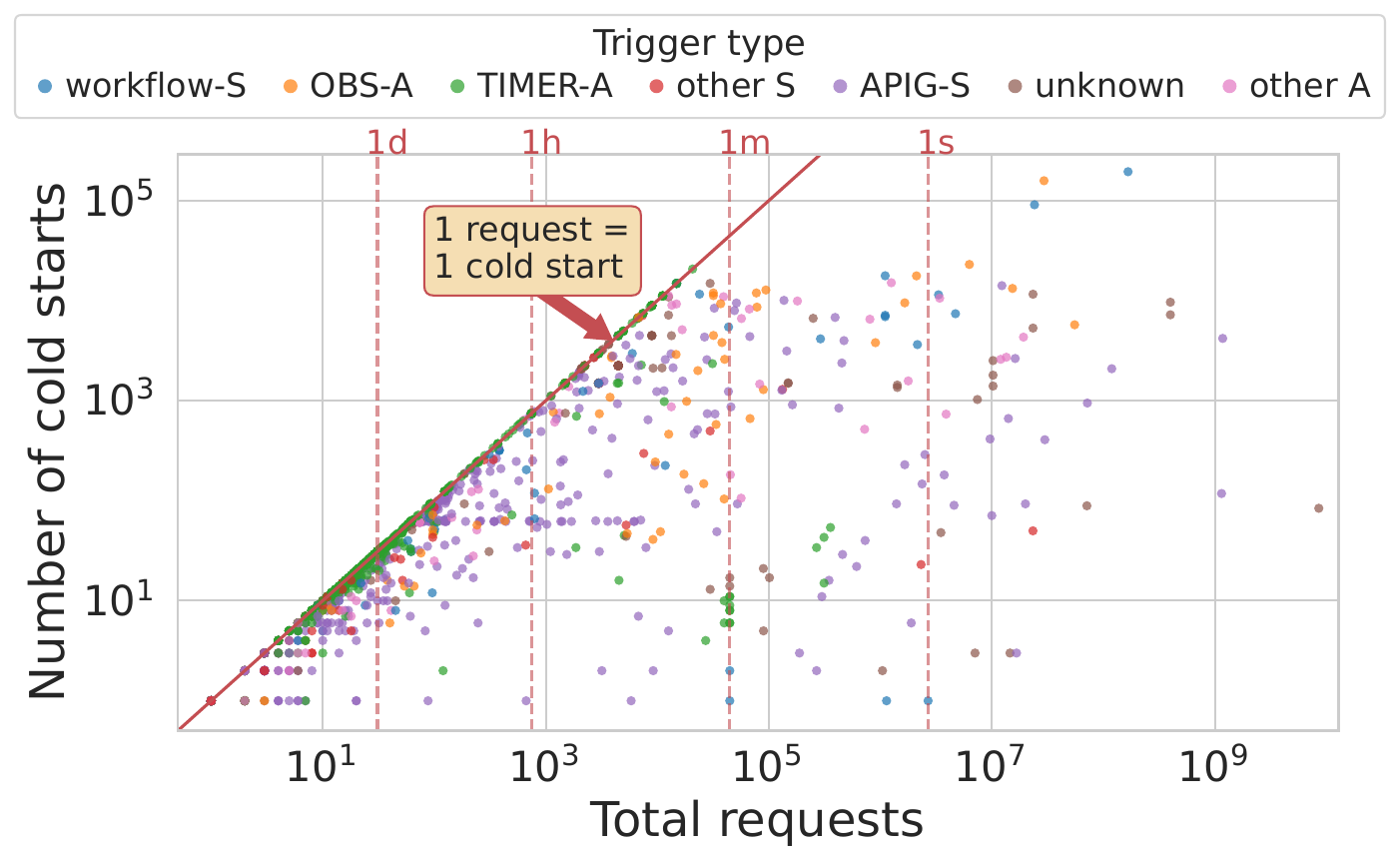}  
    \vspace{- 4 mm}
	\caption{Total number of requests per function vs number of cold starts colored by trigger type in Region 2. Each point is one function. The dashed red lines show the equivalent mean inter-arrival time.}\label{fig:total_requests_vs_number_of_cold_starts_hue_triggerType}
\end{figure}

To better understand how often cold starts occur, Figure \ref{fig:total_requests_vs_number_of_cold_starts_hue_triggerType} shows the total number of requests for each function against the total number of cold starts in Region 2, with each function colored by its trigger type. The diagonal red line represents the one-to-one case where each request causes a cold start. Most functions are invoked infrequently and therefore tend to be cold started every time they are invoked, with the majority of these being triggered by timers. Functions with more than 1 request per minute on average have fewer cold starts compared to the number of requests due to the pod keep-alive time which is set to one minute. 

\begin{tcolorbox}[title={\centering The impact of timer-triggered functions},colframe=customgrey, coltitle=black]
A significant source of cold starts is the discrepancy between timer intervals triggering functions and the pod keep-alive time. Pre-warming pods for timer-triggered functions could alleviate cold starts for timer functions. For functions with timers less frequent than the pod keep-alive time, releasing resources sooner could improve overall resource usage.
\end{tcolorbox}

\begin{figure*}
    \begin{center}{
    \captionsetup[subfigure]{margin={0.3cm,0cm}}
    \subfloat[Total cold start time.]{
       \includegraphics[height=3.03cm]{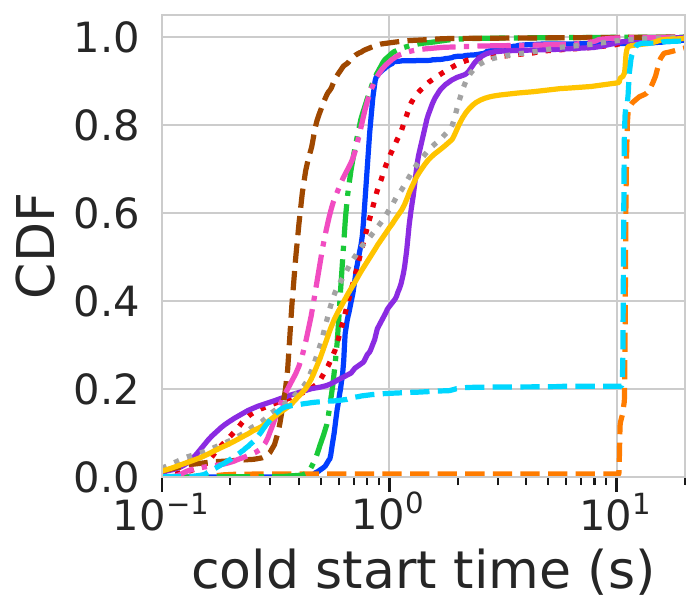}       \label{fig:cold_start_time_totalCost_by_runtime_R2} 
    }
    \captionsetup[subfigure]{margin={0cm,0cm}}
    \subfloat[Pod allocation time.]{ 
        \includegraphics[height=3.03cm]{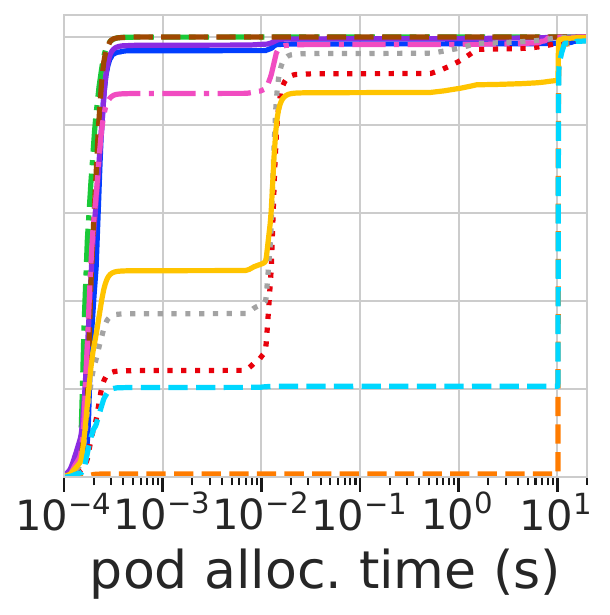}
        \label{fig:cold_start_time_getpodfrompoolCost_by_runtime_R2} 
    }   
    \subfloat[Time to deploy code.]{ 
        \includegraphics[height=3.03cm]{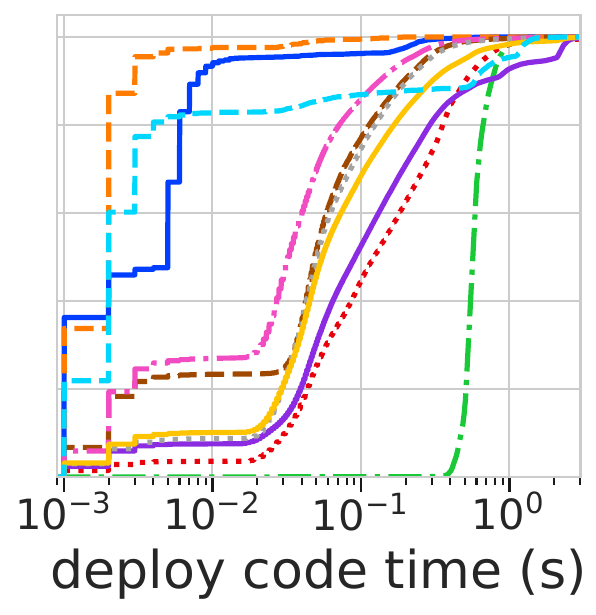}
        \label{fig:cold_start_time_deployfunctioncompressedfileCost_by_runtime_R2} 
    }
    \captionsetup[subfigure]{margin={0cm,0cm}}
    \subfloat[Deploy dependency time.]{ 
       \includegraphics[height=3.03cm]{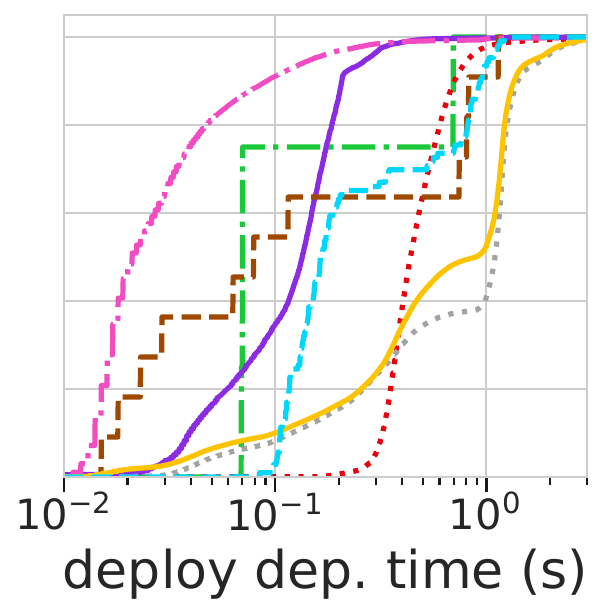}
       \label{fig:cold_start_time_deploylayersCost_by_runtime_R2} 
    }
    \captionsetup[subfigure]{margin={0cm,-0.5cm}}
    \subfloat[Scheduling time.]{      
        \includegraphics[height=3.03cm]{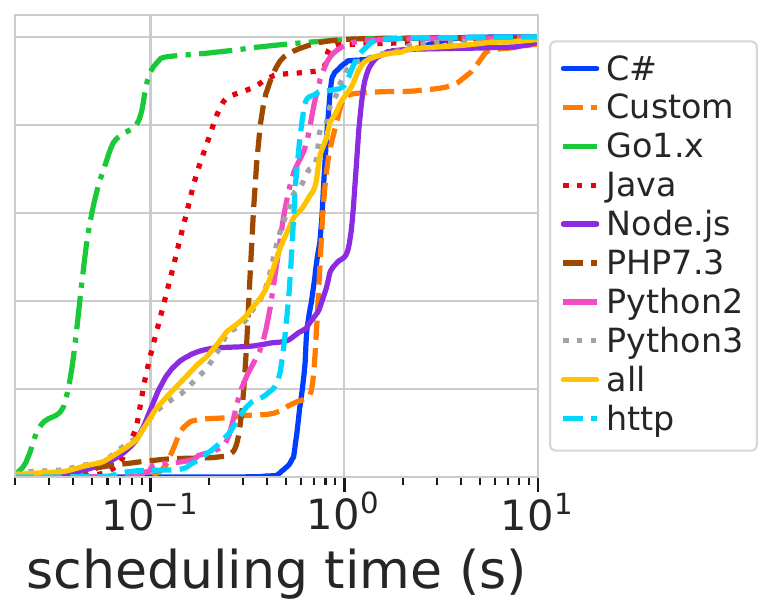}
        \label{fig:cold_start_time_remainderCost_by_runtime_R2} 
    }
    \captionsetup[subfigure]{margin={0cm,0cm}}
    }\end{center}
    \caption{Cold start time and components by runtime for Region 2.  `all' represents all cold start times combined.}
    \label{fig:cold_start_by_runtime_R2}
\end{figure*}

\begin{figure*}
    \begin{center}{
    \captionsetup[subfigure]{margin={0.3cm,0cm}}
    \subfloat[Total cold start time.]{
       \includegraphics[height=2.99cm]{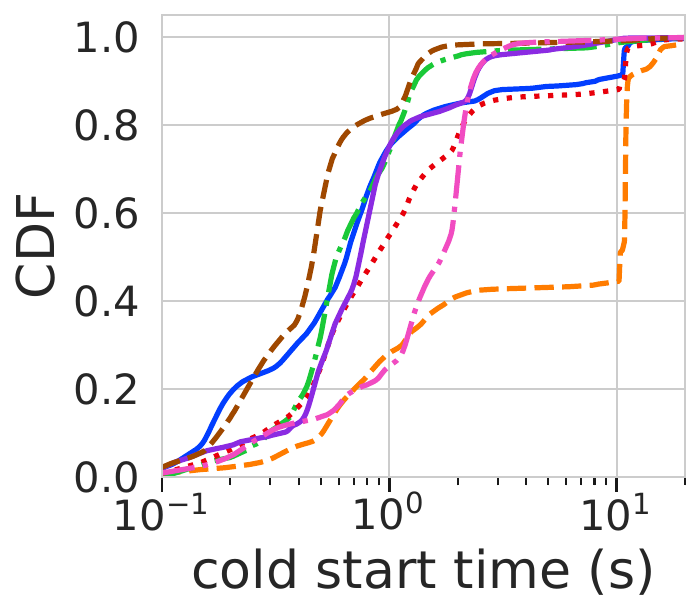}       
       \label{fig:cold_start_time_totalCost_by_triggerType} 
    }
    \captionsetup[subfigure]{margin={0cm,0cm}}
    \subfloat[Pod allocation time.]{ 
        \includegraphics[height=2.99cm]{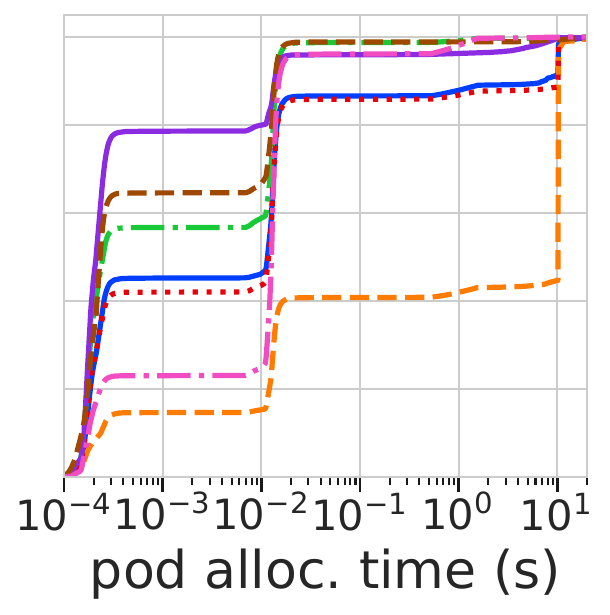}
        \label{fig:cold_start_time_getpodfrompoolCost_by_triggerType} 
    }
    \subfloat[Time to deploy code.]{ 
        \includegraphics[height=2.99cm]{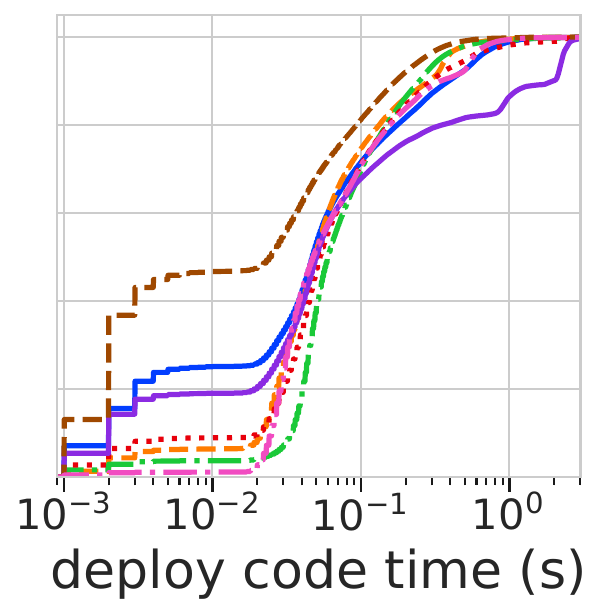}
        \label{fig:cold_start_time_deployfunctioncompressedfileCost_by_triggerType} 
    }
    \captionsetup[subfigure]{margin={0cm,0cm}}
    \subfloat[Deploy dependency time.]{ 
       \includegraphics[height=2.99cm]{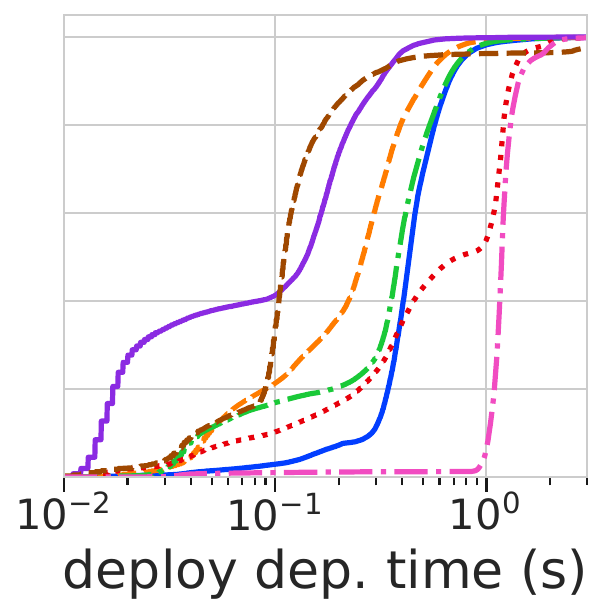}
       \label{fig:cold_start_time_deploylayersCost_by_triggerType} 
    }
    \captionsetup[subfigure]{margin={0cm,0cm}}
    \subfloat[Scheduling time.]{      
        \includegraphics[height=2.99cm]{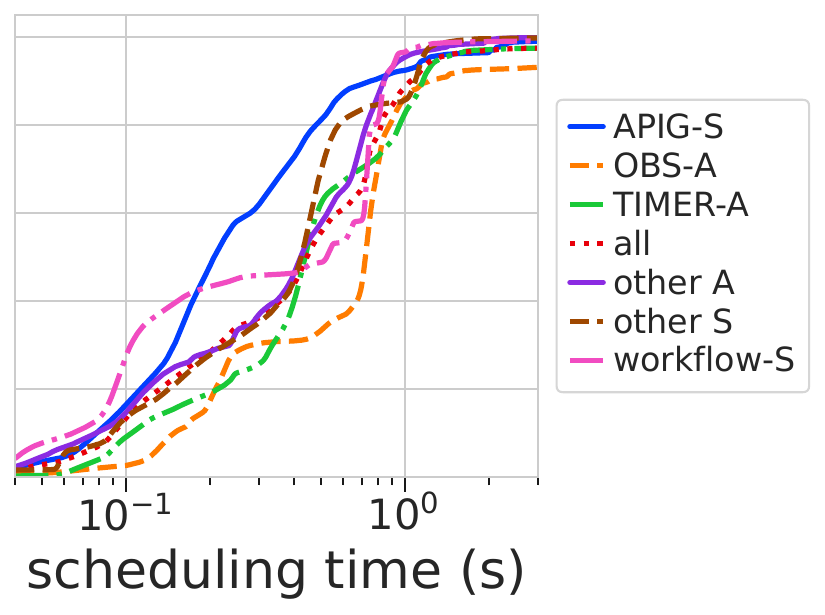}
        \label{fig:cold_start_time_remainderCost_by_triggerType} 
    }
    \captionsetup[subfigure]{margin={-0.5cm,0cm}}
    }\end{center}
    \caption{Cold start time and components by trigger type for Region 2. `all' represents all cold start times combined.}
    \label{fig:cold_start_by_triggertype_R2}
\end{figure*}

\subsection{Runtime languages and cold start time.} Previous work has found significant differences in cold start time between different runtime languages~\cite{wang2018}. Figure \ref{fig:cold_start_by_runtime_R2} shows cold start time distributions by runtime language, along with cold start component times. In all distributions, the yellow curve labelled `all' represents cold start times for all runtimes. We make the following key observations:

\begin{itemize}
    \item Cold starts in different runtimes are dominated by different components. For example, $HTTP$ cold starts are dominated by pod allocation time, while $Node.js$ cold starts are dominated by scheduling. $Go$ pods have much higher dependency and code deployment times compared to scheduling overheads.
    \item Scheduling time overheads are on average much higher than all the other overheads.
    \item For most runtimes, most cold start times are below one second with a long tail, meaning that a small proportion of cold starts takes several seconds regardless of runtime. The only exceptions are $Custom$ and $HTTP$, where the median is greater than 10 seconds.
    \item While a small fraction of running pods (see figure~\ref{fig:bar_proportions_grouped_by_runtime}), $Custom$ and $HTTP$ skew the distribution of total cold start times, increasing the tail of the combined distribution. Long cold start times from $Custom$ and $HTTP$ come from pod allocation, while other components are negligible by comparison. Slow cold starts for $Custom$ runtimes can be explained by the pod allocation process. Resource pools are not maintained for $Custom$ runtimes, so these are created from scratch every time they are required. For $HTTP$, cold starts tend to be slow since they require the start of an $HTTP$ server.
\end{itemize}

\paragraph{Trigger type and cold start time.} We now perform a similar analysis on the distribution of cold start times for different trigger types in Region 2. Figure~\ref{fig:cold_start_by_triggertype_R2} shows cold start time  and component CDFs by trigger type. Figure \ref{fig:cold_start_time_totalCost_by_triggerType} shows that functions triggered by Object Storage (OBS) tend to have a slow cold start time with a median of 10 seconds. 


It is difficult to attribute cold start times to a specific factor. However, considering the complexity of FaaS scheduling systems, as shown in Figure \ref{fig:pod_life_cycle}, it is possible to get an idea of potential bottlenecks affecting cold starts. For example, in Figure \ref{fig:cold_start_time_totalCost_by_triggerType}, the distribution for OBS has a median of 10 seconds while others have a median less than 1 second. That said, such a disparity is not necessarily caused only by the OBS trigger. Figure \ref{fig:bar_stacked_runtimes_triggerTypes} shows that the most frequently logged known trigger type for $Custom$ runtimes is OBS. In this case, the cause for the longer cold start times for OBS triggers is the fact that these functions tend to have $Custom$ runtimes, which do not have reserved resource pools and therefore require pods to be started from scratch.

\begin{tcolorbox}[title={\centering Causes of cold starts and dominant components},colframe=customgrey, coltitle=black]

There is large variability in cold start times and components driving them for different trigger types, runtimes, regions, and over time. Functions with larger resource allocations tend to have cold start times between 2 and 5 times longer compared to functions with smaller resource allocations, driven by pod allocation and code and dependency deployment. While cold starts for some trigger types, such as OBS, can be mitigated by improved networking and storage, to the best of our knowledge, there is no single solution that can reduce cold starts for all cases.
\end{tcolorbox}

\subsection{The real cost of cold starts}

\begin{figure}
    \begin{center}
        {
    \subfloat[Pod utility ratio by runtime.]{
       \includegraphics[width=0.97\linewidth]{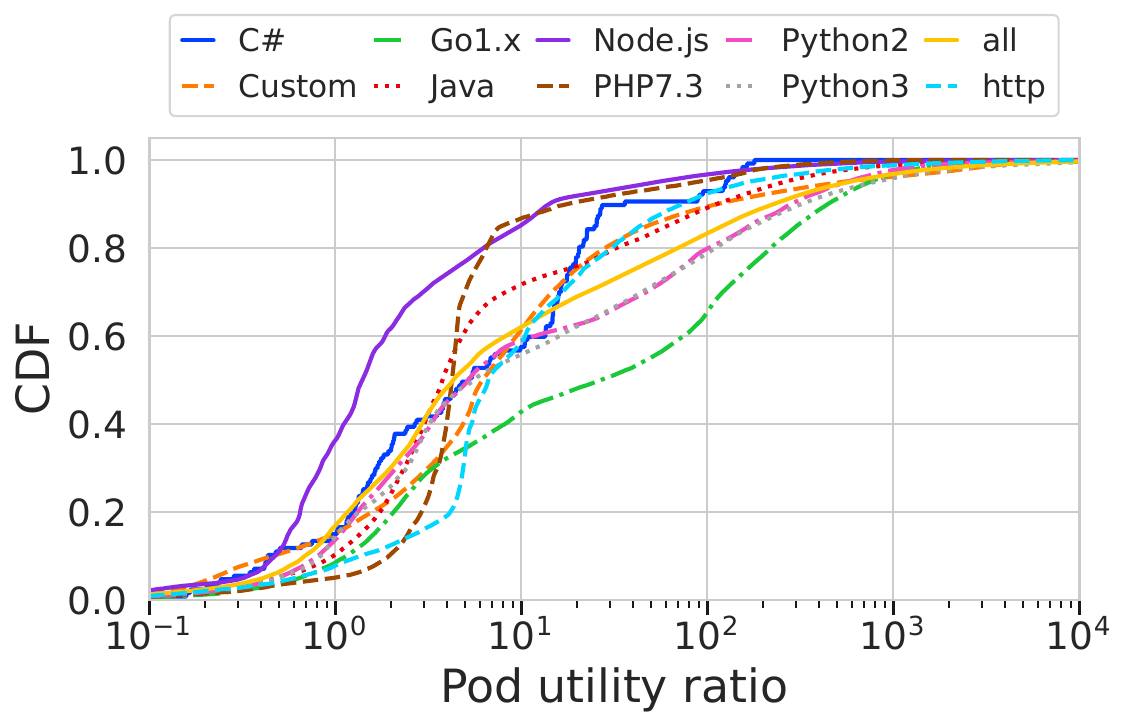}
    \label{fig:pod_lifetime_pod_lifetime_by_runtime_R2} }       
    } \end{center}
    \begin{center}
        {
    \subfloat[Pod utility ratio by trigger type.]{
       \includegraphics[width=0.97\linewidth]{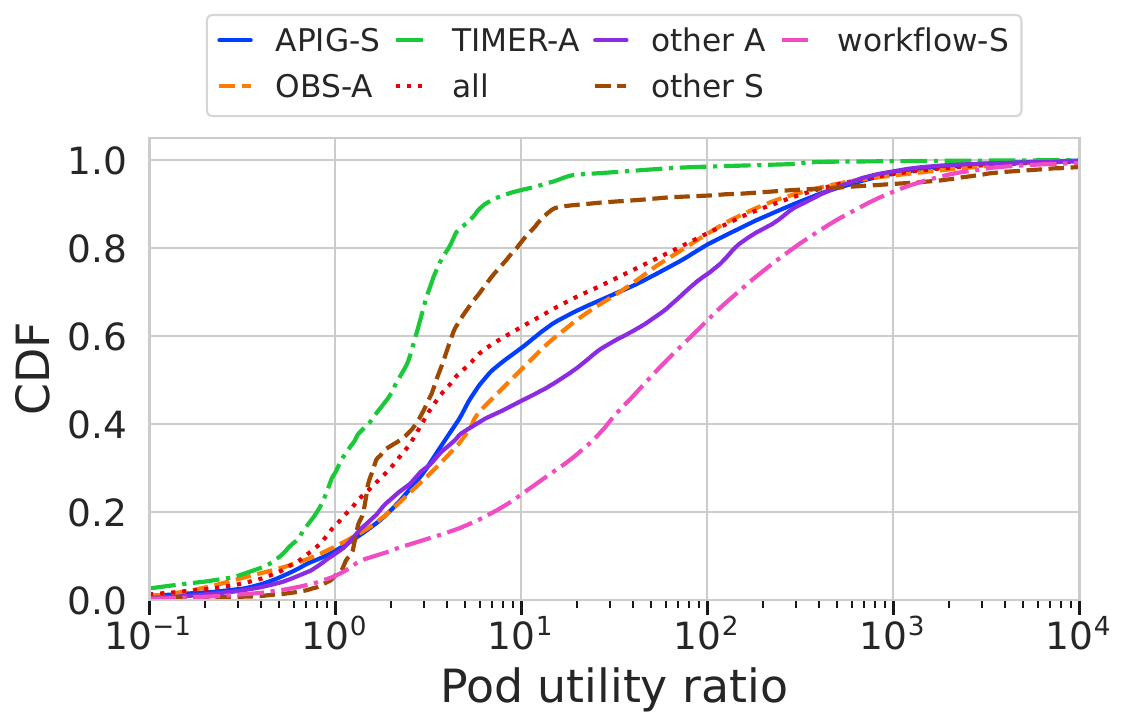}
       \label{fig:pod_lifetime_pod_lifetime_by_triggerType} 
}    }   
\end{center}
    \caption{Region 2 pod utility ratio.}
    \label{fig:pod_lifetimes_R2}
\end{figure}

Optimizing cold starts has been a popular research topic. In our work, we want to obtain a more complete understanding of cold start costs by analyzing cold starts delays relative to pod lifetime. A pod with a long cold start time is more efficient when that pod lasts longer and serves more requests than if it is deleted immediately after serving the request that spawned it. Hence, we study the ratio of a pod's useful lifetime to its cold start time. Useful pod lifetime is calculated by subtracting keep alive time (1 minute) from total pod lifetime. A 1:1 ratio or less means that the pod is used for a time less than or equal to its cold start time. A higher ratio reflects a greater utility to the function and the system. We hence call this the `utility ratio'. 

Figure~\ref{fig:pod_lifetime_pod_lifetime_by_runtime_R2} shows utility ratios  per runtime, while Figure~\ref{fig:pod_lifetime_pod_lifetime_by_triggerType} shows utility ratios per trigger type. We see that 20\% of all pods have a pod utility ratio less than 1, meaning that their useful lifetime is less than their cold start time. Median utility ratio is approximately 4:1, meaning that 50\% of pods last for 4 times their cold start time.
In Figure~\ref{fig:pod_lifetime_pod_lifetime_by_runtime_R2}, we see that a significant number of cold starts from $Node.js$, about 40\%, had a ratio lower than 1. $Node.js$ cold starts represent more than 10\% of all the cold-starts in our system (see Figure~\ref{fig:bar_proportions_grouped_by_runtime}). $Node.js$ is also the third slowest runtime in our system (see Figure~\ref{fig:cold_start_by_runtime_R2}). The runtimes with the second worst utility ratio are $PHP7.3$ and $Java$, both of which have at least 70\% of their cold starts having a utility ratio lower than 10. The highest utility ratio runtime is $Go1.x$ where 35\% of pods have a utility ratio greater than 100. Notably, $Custom$ and $HTTP$ runtimes have a utility ratio better than several default runtimes, which have much shorter cold start times. Hence, pod utility ratio can be used to obtain a different perspective on the cost of cold starts, particularly for pods with long cold start times, to see how long those pods remain in the system. 

Figure~\ref{fig:pod_lifetime_pod_lifetime_by_triggerType} shows that timers have the lowest utility ratios, which corresponds with most timer functions being cold started as discussed in previous sections. Timer functions, despite having some of the shortest cold start times, have a low pod utility ratio compared to other trigger types. Meanwhile, workflow-S, which has longer cold start times, tends to have higher utility ratios compared to other trigger types.

\begin{tcolorbox}[title={\centering Utility ratios},colframe=customgrey, coltitle=black]
Utility ratios help obtain a more complete picture of the cost of cold starts beyond cold start time by accounting for how long these pods remain useful in the system. We find that runtimes and trigger types with long cold start times tend to have higher pod utility ratios.
\end{tcolorbox}

\section{Discussion and open problems}\label{section:discussion_and_open_problems}

Serverless systems research has gained momentum, with many projects aimed at reducing cold starts. Recently, \citet{liu2023gap} identified five research problems for future work on serverless systems. In this section, we leverage our analysis from previous sections to identify several areas important for serverless system optimization.

\paragraph{Cross-region workload scheduling.} The average latency between data centers in developed regions is relatively small, in the orders of tens to a few hundred milliseconds~\cite{zeng2022congestion, 10188700} with newer cross data center networking solutions expected to reduce this latency further~\cite{dukic2020beyond}. In our analysis, the most popular regions consistently have much longer average, median, and tail cold-start times. While some of these differences can be attributed to differences in networking, our analysis shows significant differences in cold start time, CPU usage, and memory usage exist between regions, and less congested regions may offer cheaper and faster options for running workloads. It is therefore important for the research community to study cross-region serverless platforms and load-balancing. Cross-region scheduling could assume the form of a global or fleet-wide control plane accounting for peak time shifts, overall system load, latency from the user to the data center, and resources in different regions.

\paragraph{Scheduling delays are significant.} \citet{liu2023gap} discussed how a large component of cold starts comes from scheduling delays. Our analysis shows that in our five data centers, except for a brief period in Region 1, scheduling delays are either the most or the second most significant component in cold start times. While one can think of optimizations to reduce these delays, it might be that a fundamental redesign of serverless platforms would be required. Currently, most serverless platforms run in a multi-layered stack, e.g. on top of Kubernetes, Ray, firecracker, KVM, or a combination of these systems. This adds multiple layers of complexity during a cold start, including setup, configurations, initialization, and scheduling. We believe that novel serverless platforms that reduce layered complexity would result in more optimized serverless performance.

\paragraph{Synchronous vs. asynchronous calls.} In our workloads, roughly 60\% of functions are asynchronous, in line with \citet{liu2023gap}. However, when studying the time series of running pods in Figure~\ref{fig:triggerTypeTS} and the bar chart in Figure~\ref{fig:bar_proportions_grouped_by_triggerType_invocationType}, we see that cold starts of functions with synchronous triggers sometimes dominate the total number of cold starts. This points to the fact that optimizations should consider both asynchronous and synchronous functions. Such optimizations can include peak shaving of asynchronous functions if these are not latency critical, starting them for example when there are much fewer synchronous functions.

\paragraph{Predicting cold starts.} In Figure \ref{fig:peak_trough_ratio}, we show that cold starts may be caused by functions with small or large peak-to-trough ratios. These tend to be low-request functions that are cold started every time they are invoked, such as those along the diagonal line in Figure~\ref{fig:total_requests_vs_number_of_cold_starts_hue_triggerType}, or high-request functions with larger peak-to-trough ratios. These two types of functions may require different solutions for predicting resource allocation needs, with the highly requested ones potentially benefiting from periodic time series predictions~\cite{Joosen_2023}. Functions running on timer triggers could be pre-warmed before their next invocation. Similarly, for functions running on timers less frequent than 1 minute, a keep alive time of 1 minute is unnecessary and wasteful. Cloud providers may consider a dynamic keep-alive time for such functions.


\paragraph{Workflow function calls can be predicted using previous function calls.} As our analysis show, a significant number of cold starts occur due to synchronous workflow functions which can be predicted using function calls earlier in the chain. Resources for downstream functions could be allocated based on the invocations (and maybe even resource usage) of function calls that will invoke it later. This may alleviate the relatively slow cold start time of workflow-triggered functions. Currently, workflows account for 20\% of cold starts. Additionally, the synchronous nature of these requests demands a strict latency SLO, which can be vastly improved with predictive autoscaling. Collection and analysis of function call chains may show opportunities for improvement.

\paragraph{Concurrency adjustment.} Each function has a user-set concurrency value that determines how many function requests can be executed at the same time. For many functions, the resource utilization can be improved by increasing concurrency as long as the total execution time remains acceptable. This is especially useful given the strong oscillations in some asynchronous triggers, such as OBS.

\paragraph{Resource pool prediction.} Our system maintains pools of inactive pods to be used as demanded by user traffic. If demand exceeds the capacity of the resource pool, a pod will be started from scratch, causing significant delays. However, keeping an unnecessarily large number of pods in the resource pool may be wasteful. Due to predictable time-varying patterns of various pod configurations, such as pods of different resource configurations, it may be possible to predict the required number of reserved pods so that user demand is met without unnecessary overallocation. This differs from function invocation prediction~\cite{Joosen_2023, foldformer_2023, shahrad2020serverless}, which requires additional steps from number of invocations to a scaling decision, and instead directly predicts required resources. 

\section{Related work}

Analyzing and characterizing cloud and systems workloads has enabled the systems community to perform research on improving computer systems performance.
Google released multiple traces from their internal Borg system~\cite{verma2015large,clusterdata:Tirmazi2020,clusterdata:Wilkes2020a,clusterdata:Reiss2012}. This data has been extensively analyzed by the research community over the years~\cite{clusterdata:jajooSLearn2022,clusterdata:Tirmazi2020,clusterdata:Sebastio2018,clusterdata:Reiss2012b}. Since then, there have been other data releases from multiple cloud providers for different internal systems, including Alibaba~\cite{guo2019limits,liu2018elasticity,weng2023beware, weng2022mlaas}, Azure~\cite{shahrad2020serverless,ResourceCentral2017, hadary2020protean}, and Huawei~\cite{Joosen_2023}.
Earlier research has shown that one of the main bottlenecks of serverless systems is cold starts, where the system does not have sufficient resources to process an incoming request and must start a new pod from scratch~\cite{Joosen_2023, weng2022mlaas}. There have been significant efforts to optimize cold starts, but few of these are informed by insights from production data. 

Prior work with specific mention of cold start statistics~\cite{Joosen_2023, shahrad2020serverless} tends to offer high-level metrics from a single region with little discussion of components and the effect of factors such as runtime language, resource allocation, and trigger type on the number of cold starts and their component times. Our work analyzes granular event-level metrics with detailed component times of cold starts from five regions, and examines the effect of function characteristics such as resource allocation, runtime language, and trigger type. 

Cold starts have previously been found to be affected by function memory allocation, runtime language, and network latency~\cite{wang2018}. Reducing cold starts using novel techniques to calculate the keep-alive time of a container is one active research topic~\cite{fuerst2021faascache}. Another interesting research direction is optimizing container deployments for serverless functions~\cite{oakes2018sock,FaaSNet_2021}. Our work enables systems researchers to better understand some of the bottlenecks in production serverless systems including some of the root causes of cold starts.

\section{Conclusion}

This paper conducts an analysis of a multi-region production serverless cloud platform, focusing on factors affecting cold start times and their components. We have examined these factors in the context of long-term, evolving workloads as well as a week long holiday period in our month long dataset. 

Our study reveals significant variations in function execution time, resource usage, and cold start time between regions, which may point to benefits of cross-region load balancing to reduce overall latency and cost. In all of our regions, cold start time is positively correlated with the number of cold starts. The dominant component of cold start time tends to vary between regions, which may point to workload differences or bottlenecks in different parts of the architecture. In all regions, the number of cold starts tends to decrease during the holiday period, with a `catch-up' period afterwards where the number of cold starts and cold start time increase significantly. Additionally, cold start time for larger resource allocations tends to be longer than for smaller resource allocations, with pod allocation and deployment time for code and dependencies being contributors. 

Furthermore, we have determined that significant numbers of cold starts come from several types of functions, such as low-request timer-triggered functions or high-request functions with large peak-to-trough ratios that require frequent autoscaling. We find that factors such as trigger type and runtime language affect the number and duration of cold starts differently. For example, pods using $Custom$ runtimes experience significantly longer cold starts than those using default runtimes due to the absence of a reserved pool, with pod allocation time accounting for nearly the entire cold start duration. We introduce pod utility ratio, which can be used to measure a pod's usefulness by computing the ratio between a pod's useful lifetime (excluding keep-alive time) and its cold start time. We find that, in some cases, pod configurations with long cold start times tend to last longer, such as for $Custom$ runtimes. Finally, we leverage our analysis to identify several bottlenecks in serverless public cloud platforms that contribute to cold starts and highlight opportunities for future research to address these issues and improve performance. 

\section*{Acknowledgments}
We would like to thank the YuanRong team at Huawei for their valuable collaboration and helping to collect the data. We would also like to thank Wei Wei for his feedback. Lastly, we would like to thank the reviewers at EuroSys for their insightful comments and Laiping Zhao for shepherding.



\bibliographystyle{ACM-Reference-Format}
\bibliography{eurosys2023}


\begin{thebibliography}{46}


\ifx \showCODEN    \undefined \def \showCODEN     #1{\unskip}     \fi
\ifx \showDOI      \undefined \def \showDOI       #1{#1}\fi
\ifx \showISBNx    \undefined \def \showISBNx     #1{\unskip}     \fi
\ifx \showISBNxiii \undefined \def \showISBNxiii  #1{\unskip}     \fi
\ifx \showISSN     \undefined \def \showISSN      #1{\unskip}     \fi
\ifx \showLCCN     \undefined \def \showLCCN      #1{\unskip}     \fi
\ifx \shownote     \undefined \def \shownote      #1{#1}          \fi
\ifx \showarticletitle \undefined \def \showarticletitle #1{#1}   \fi
\ifx \showURL      \undefined \def \showURL       {\relax}        \fi
\providecommand\bibfield[2]{#2}
\providecommand\bibinfo[2]{#2}
\providecommand\natexlab[1]{#1}
\providecommand\showeprint[2][]{arXiv:#2}

\bibitem[Agache et~al\mbox{.}(2020)]%
        {agache2020firecracker}
\bibfield{author}{\bibinfo{person}{Alexandru Agache}, \bibinfo{person}{Marc Brooker}, \bibinfo{person}{Alexandra Iordache}, \bibinfo{person}{Anthony Liguori}, \bibinfo{person}{Rolf Neugebauer}, \bibinfo{person}{Phil Piwonka}, {and} \bibinfo{person}{Diana-Maria Popa}.} \bibinfo{year}{2020}\natexlab{}.
\newblock \showarticletitle{Firecracker: Lightweight virtualization for serverless applications}. In \bibinfo{booktitle}{\emph{17th USENIX symposium on networked systems design and implementation (NSDI 20)}}. \bibinfo{pages}{419--434}.
\newblock


\bibitem[Anderson et~al\mbox{.}(2023)]%
        {anderson2023treehouse}
\bibfield{author}{\bibinfo{person}{Thomas Anderson}, \bibinfo{person}{Adam Belay}, \bibinfo{person}{Mosharaf Chowdhury}, \bibinfo{person}{Asaf Cidon}, {and} \bibinfo{person}{Irene Zhang}.} \bibinfo{year}{2023}\natexlab{}.
\newblock \showarticletitle{Treehouse: A case for carbon-aware datacenter software}.
\newblock \bibinfo{journal}{\emph{ACM SIGENERGY Energy Informatics Review}} \bibinfo{volume}{3}, \bibinfo{number}{3} (\bibinfo{year}{2023}), \bibinfo{pages}{64--70}.
\newblock


\bibitem[Bashir et~al\mbox{.}(2021)]%
        {bashir2021enabling}
\bibfield{author}{\bibinfo{person}{Noman Bashir}, \bibinfo{person}{Tian Guo}, \bibinfo{person}{Mohammad Hajiesmaili}, \bibinfo{person}{David Irwin}, \bibinfo{person}{Prashant Shenoy}, \bibinfo{person}{Ramesh Sitaraman}, \bibinfo{person}{Abel Souza}, {and} \bibinfo{person}{Adam Wierman}.} \bibinfo{year}{2021}\natexlab{}.
\newblock \showarticletitle{Enabling sustainable clouds: The case for virtualizing the energy system}. In \bibinfo{booktitle}{\emph{Proceedings of the ACM Symposium on Cloud Computing}}. \bibinfo{pages}{350--358}.
\newblock


\bibitem[Bodik et~al\mbox{.}(2010)]%
        {bodik2010characterizing}
\bibfield{author}{\bibinfo{person}{Peter Bodik}, \bibinfo{person}{Armando Fox}, \bibinfo{person}{Michael~J Franklin}, \bibinfo{person}{Michael~I Jordan}, {and} \bibinfo{person}{David~A Patterson}.} \bibinfo{year}{2010}\natexlab{}.
\newblock \showarticletitle{Characterizing, modeling, and generating workload spikes for stateful services}. In \bibinfo{booktitle}{\emph{Proceedings of the 1st ACM symposium on Cloud computing}}. \bibinfo{pages}{241--252}.
\newblock


\bibitem[Chen et~al\mbox{.}(2022)]%
        {chen2022starlight}
\bibfield{author}{\bibinfo{person}{Jun~Lin Chen}, \bibinfo{person}{Daniyal Liaqat}, \bibinfo{person}{Moshe Gabel}, {and} \bibinfo{person}{Eyal de Lara}.} \bibinfo{year}{2022}\natexlab{}.
\newblock \showarticletitle{Starlight: Fast container provisioning on the edge and over the WAN}. In \bibinfo{booktitle}{\emph{19th USENIX Symposium on Networked Systems Design and Implementation (NSDI 22)}}. \bibinfo{pages}{35--50}.
\newblock


\bibitem[Chen et~al\mbox{.}(2024)]%
        {chen2024yuanrong}
\bibfield{author}{\bibinfo{person}{Qiong Chen}, \bibinfo{person}{Jianmin Qian}, \bibinfo{person}{Yulin Che}, \bibinfo{person}{Ziqi Lin}, \bibinfo{person}{Jianfeng Wang}, \bibinfo{person}{Jie Zhou}, \bibinfo{person}{Licheng Song}, \bibinfo{person}{Yi Liang}, \bibinfo{person}{Jie Wu}, \bibinfo{person}{Wei Zheng}, {et~al\mbox{.}}} \bibinfo{year}{2024}\natexlab{}.
\newblock \showarticletitle{YuanRong: A production general-purpose serverless system for distributed applications in the cloud}. In \bibinfo{booktitle}{\emph{Proceedings of the ACM SIGCOMM 2024 Conference}}. \bibinfo{pages}{843--859}.
\newblock


\bibitem[Cortez et~al\mbox{.}(2017)]%
        {ResourceCentral2017}
\bibfield{author}{\bibinfo{person}{Eli Cortez}, \bibinfo{person}{Anand Bonde}, \bibinfo{person}{Alexandre Muzio}, \bibinfo{person}{Mark Russinovich}, \bibinfo{person}{Marcus Fontoura}, {and} \bibinfo{person}{Ricardo Bianchini}.} \bibinfo{year}{2017}\natexlab{}.
\newblock \showarticletitle{Resource Central: Understanding and Predicting Workloads for Improved Resource Management in Large Cloud Platforms}. In \bibinfo{booktitle}{\emph{Proceedings of the 26th Symposium on Operating Systems Principles}} (Shanghai, China) \emph{(\bibinfo{series}{SOSP '17})}. \bibinfo{publisher}{Association for Computing Machinery}, \bibinfo{address}{New York, NY, USA}, \bibinfo{pages}{153–167}.
\newblock
\showISBNx{9781450350853}
\urldef\tempurl%
\url{https://doi.org/10.1145/3132747.3132772}
\showDOI{\tempurl}


\bibitem[Darlow et~al\mbox{.}(2023)]%
        {foldformer_2023}
\bibfield{author}{\bibinfo{person}{Luke~Nicholas Darlow}, \bibinfo{person}{Artjom Joosen}, \bibinfo{person}{Martin Asenov}, \bibinfo{person}{Qiwen Deng}, \bibinfo{person}{Jianfeng Wang}, {and} \bibinfo{person}{Adam Barker}.} \bibinfo{year}{2023}\natexlab{}.
\newblock \showarticletitle{FoldFormer: sequence folding and seasonal attention for fine-grained long-term FaaS forecasting}. In \bibinfo{booktitle}{\emph{Proceedings of the 3rd Workshop on Machine Learning and Systems}} (Rome, Italy) \emph{(\bibinfo{series}{EuroMLSys '23})}. \bibinfo{publisher}{Association for Computing Machinery}, \bibinfo{address}{New York, NY, USA}, \bibinfo{pages}{71–77}.
\newblock
\showISBNx{9798400700842}
\urldef\tempurl%
\url{https://doi.org/10.1145/3578356.3592582}
\showDOI{\tempurl}


\bibitem[Dukic et~al\mbox{.}(2020)]%
        {dukic2020beyond}
\bibfield{author}{\bibinfo{person}{Vojislav Dukic}, \bibinfo{person}{Ginni Khanna}, \bibinfo{person}{Christos Gkantsidis}, \bibinfo{person}{Thomas Karagiannis}, \bibinfo{person}{Francesca Parmigiani}, \bibinfo{person}{Ankit Singla}, \bibinfo{person}{Mark Filer}, \bibinfo{person}{Jeffrey~L Cox}, \bibinfo{person}{Anna Ptasznik}, \bibinfo{person}{Nick Harland}, {et~al\mbox{.}}} \bibinfo{year}{2020}\natexlab{}.
\newblock \showarticletitle{Beyond the mega-data center: Networking multi-data center regions}. In \bibinfo{booktitle}{\emph{Proceedings of the Annual conference of the ACM Special Interest Group on Data Communication on the applications, technologies, architectures, and protocols for computer communication}}. \bibinfo{pages}{765--781}.
\newblock


\bibitem[Fuerst and Sharma(2021)]%
        {fuerst2021faascache}
\bibfield{author}{\bibinfo{person}{Alexander Fuerst} {and} \bibinfo{person}{Prateek Sharma}.} \bibinfo{year}{2021}\natexlab{}.
\newblock \showarticletitle{FaasCache: keeping serverless computing alive with greedy-dual caching}. In \bibinfo{booktitle}{\emph{Proceedings of the 26th ACM International Conference on Architectural Support for Programming Languages and Operating Systems}}. \bibinfo{pages}{386--400}.
\newblock


\bibitem[Geng et~al\mbox{.}(2023)]%
        {10188700}
\bibfield{author}{\bibinfo{person}{Yantao Geng}, \bibinfo{person}{Han Zhang}, \bibinfo{person}{Xingang Shi}, \bibinfo{person}{Jilong Wang}, \bibinfo{person}{Xia Yin}, \bibinfo{person}{Dongbiao He}, {and} \bibinfo{person}{Yahui Li}.} \bibinfo{year}{2023}\natexlab{}.
\newblock \showarticletitle{Delay Based Congestion Control for Cross-Datacenter Networks}. In \bibinfo{booktitle}{\emph{2023 IEEE/ACM 31st International Symposium on Quality of Service (IWQoS)}}. \bibinfo{pages}{1--4}.
\newblock
\urldef\tempurl%
\url{https://doi.org/10.1109/IWQoS57198.2023.10188700}
\showDOI{\tempurl}


\bibitem[Guo et~al\mbox{.}(2019)]%
        {guo2019limits}
\bibfield{author}{\bibinfo{person}{Jing Guo}, \bibinfo{person}{Zihao Chang}, \bibinfo{person}{Sa Wang}, \bibinfo{person}{Haiyang Ding}, \bibinfo{person}{Yihui Feng}, \bibinfo{person}{Liang Mao}, {and} \bibinfo{person}{Yungang Bao}.} \bibinfo{year}{2019}\natexlab{}.
\newblock \showarticletitle{Who limits the resource efficiency of my datacenter: An analysis of alibaba datacenter traces}. In \bibinfo{booktitle}{\emph{Proceedings of the international symposium on quality of service}}. \bibinfo{pages}{1--10}.
\newblock


\bibitem[Hadary et~al\mbox{.}(2020)]%
        {hadary2020protean}
\bibfield{author}{\bibinfo{person}{Ori Hadary}, \bibinfo{person}{Luke Marshall}, \bibinfo{person}{Ishai Menache}, \bibinfo{person}{Abhisek Pan}, \bibinfo{person}{Esaias~E Greeff}, \bibinfo{person}{David Dion}, \bibinfo{person}{Star Dorminey}, \bibinfo{person}{Shailesh Joshi}, \bibinfo{person}{Yang Chen}, \bibinfo{person}{Mark Russinovich}, {et~al\mbox{.}}} \bibinfo{year}{2020}\natexlab{}.
\newblock \showarticletitle{Protean: VM allocation service at scale}. In \bibinfo{booktitle}{\emph{14th USENIX Symposium on Operating Systems Design and Implementation (OSDI 20)}}. \bibinfo{pages}{845--861}.
\newblock


\bibitem[Hanafy et~al\mbox{.}(2023)]%
        {hanafy2023carbonscaler}
\bibfield{author}{\bibinfo{person}{Walid~A Hanafy}, \bibinfo{person}{Qianlin Liang}, \bibinfo{person}{Noman Bashir}, \bibinfo{person}{David Irwin}, {and} \bibinfo{person}{Prashant Shenoy}.} \bibinfo{year}{2023}\natexlab{}.
\newblock \showarticletitle{CarbonScaler: Leveraging Cloud Workload Elasticity for Optimizing Carbon-Efficiency}.
\newblock \bibinfo{journal}{\emph{Proceedings of the ACM on Measurement and Analysis of Computing Systems}} \bibinfo{volume}{7}, \bibinfo{number}{3} (\bibinfo{year}{2023}), \bibinfo{pages}{1--28}.
\newblock


\bibitem[Huang et~al\mbox{.}(2022)]%
        {huang2022metastable}
\bibfield{author}{\bibinfo{person}{Lexiang Huang}, \bibinfo{person}{Matthew Magnusson}, \bibinfo{person}{Abishek~Bangalore Muralikrishna}, \bibinfo{person}{Salman Estyak}, \bibinfo{person}{Rebecca Isaacs}, \bibinfo{person}{Abutalib Aghayev}, \bibinfo{person}{Timothy Zhu}, {and} \bibinfo{person}{Aleksey Charapko}.} \bibinfo{year}{2022}\natexlab{}.
\newblock \showarticletitle{Metastable failures in the wild}. In \bibinfo{booktitle}{\emph{16th USENIX Symposium on Operating Systems Design and Implementation (OSDI 22)}}. \bibinfo{pages}{73--90}.
\newblock


\bibitem[iguazio(2024)]%
        {Nuclio}
\bibfield{author}{\bibinfo{person}{iguazio}.} \bibinfo{year}{2024}\natexlab{}.
\newblock \bibinfo{title}{Nuclio}.
\newblock
\newblock
\urldef\tempurl%
\url{https://nuclio.io/}
\showURL{%
\tempurl}
\newblock
\shownote{Accessed: May, 2024}.


\bibitem[Jajoo et~al\mbox{.}(2022)]%
        {clusterdata:jajooSLearn2022}
\bibfield{author}{\bibinfo{person}{Akshay Jajoo}, \bibinfo{person}{Y.~Charlie Hu}, \bibinfo{person}{Xiaojun Lin}, {and} \bibinfo{person}{Nan Deng}.} \bibinfo{year}{2022}\natexlab{}.
\newblock \showarticletitle{A Case for Task Sampling based Learning for Cluster Job Scheduling}. In \bibinfo{booktitle}{\emph{19th USENIX Symposium on Networked Systems Design and Implementation (NSDI 22)}}. \bibinfo{publisher}{USENIX Association}, \bibinfo{address}{Renton, WA, USA}.
\newblock
\urldef\tempurl%
\url{https://www.usenix.org/conference/nsdi22/presentation/jajoo}
\showURL{%
\tempurl}


\bibitem[Joosen et~al\mbox{.}(2023)]%
        {Joosen_2023}
\bibfield{author}{\bibinfo{person}{Artjom Joosen}, \bibinfo{person}{Ahmed Hassan}, \bibinfo{person}{Martin Asenov}, \bibinfo{person}{Rajkarn Singh}, \bibinfo{person}{Luke Darlow}, \bibinfo{person}{Jianfeng Wang}, {and} \bibinfo{person}{Adam Barker}.} \bibinfo{year}{2023}\natexlab{}.
\newblock \showarticletitle{How Does It Function? Characterizing Long-term Trends in Production Serverless Workloads}. In \bibinfo{booktitle}{\emph{Proceedings of the 2023 ACM Symposium on Cloud Computing}} \emph{(\bibinfo{series}{SoCC ’23})}. \bibinfo{publisher}{ACM}.
\newblock
\urldef\tempurl%
\url{https://doi.org/10.1145/3620678.3624783}
\showDOI{\tempurl}


\bibitem[{Knative Team}(2024)]%
        {Knative}
\bibfield{author}{\bibinfo{person}{{Knative Team}}.} \bibinfo{year}{2024}\natexlab{}.
\newblock \bibinfo{title}{Knative: Concepts}.
\newblock
\newblock
\urldef\tempurl%
\url{https://knative.dev/docs/concepts/}
\showURL{%
\tempurl}
\newblock
\shownote{Accessed: May, 2024}.


\bibitem[Kızılersü et~al\mbox{.}(2018)]%
        {weibull}
\bibfield{author}{\bibinfo{person}{Ayşe Kızılersü}, \bibinfo{person}{Markus Kreer}, {and} \bibinfo{person}{Anthony~W. Thomas}.} \bibinfo{year}{2018}\natexlab{}.
\newblock \showarticletitle{{The Weibull Distribution}}.
\newblock \bibinfo{journal}{\emph{Significance}} \bibinfo{volume}{15}, \bibinfo{number}{2} (\bibinfo{date}{04} \bibinfo{year}{2018}), \bibinfo{pages}{10--11}.
\newblock
\showISSN{1740-9705}
\urldef\tempurl%
\url{https://doi.org/10.1111/j.1740-9713.2018.01123.x}
\showDOI{\tempurl}
\showeprint{https://academic.oup.com/jrssig/article-pdf/15/2/10/49185942/sign\_15\_2\_10.pdf}


\bibitem[Li(2023)]%
        {li2023modernization}
\bibfield{author}{\bibinfo{person}{Feifei Li}.} \bibinfo{year}{2023}\natexlab{}.
\newblock \showarticletitle{Modernization of databases in the cloud era: Building databases that run like Legos}.
\newblock \bibinfo{journal}{\emph{Proceedings of the VLDB Endowment}} \bibinfo{volume}{16}, \bibinfo{number}{12} (\bibinfo{year}{2023}), \bibinfo{pages}{4140--4151}.
\newblock


\bibitem[Li et~al\mbox{.}(2021)]%
        {li2021analyzing}
\bibfield{author}{\bibinfo{person}{Junfeng Li}, \bibinfo{person}{Sameer~G Kulkarni}, \bibinfo{person}{KK Ramakrishnan}, {and} \bibinfo{person}{Dan Li}.} \bibinfo{year}{2021}\natexlab{}.
\newblock \showarticletitle{Analyzing open-source serverless platforms: Characteristics and performance}.
\newblock \bibinfo{journal}{\emph{arXiv preprint arXiv:2106.03601}} (\bibinfo{year}{2021}).
\newblock


\bibitem[Li et~al\mbox{.}(2022)]%
        {li2022help}
\bibfield{author}{\bibinfo{person}{Zijun Li}, \bibinfo{person}{Linsong Guo}, \bibinfo{person}{Quan Chen}, \bibinfo{person}{Jiagan Cheng}, \bibinfo{person}{Chuhao Xu}, \bibinfo{person}{Deze Zeng}, \bibinfo{person}{Zhuo Song}, \bibinfo{person}{Tao Ma}, \bibinfo{person}{Yong Yang}, \bibinfo{person}{Chao Li}, {et~al\mbox{.}}} \bibinfo{year}{2022}\natexlab{}.
\newblock \showarticletitle{Help rather than recycle: Alleviating cold startup in serverless computing through Inter-Function container sharing}. In \bibinfo{booktitle}{\emph{2022 USENIX Annual Technical Conference (USENIX ATC 22)}}. \bibinfo{pages}{69--84}.
\newblock


\bibitem[Liu et~al\mbox{.}(2023b)]%
        {liu2023doing}
\bibfield{author}{\bibinfo{person}{David~H Liu}, \bibinfo{person}{Amit Levy}, \bibinfo{person}{Shadi Noghabi}, {and} \bibinfo{person}{Sebastian Burckhardt}.} \bibinfo{year}{2023}\natexlab{b}.
\newblock \showarticletitle{Doing more with less: orchestrating serverless applications without an orchestrator}. In \bibinfo{booktitle}{\emph{20th USENIX Symposium on Networked Systems Design and Implementation (NSDI 23)}}. \bibinfo{pages}{1505--1519}.
\newblock


\bibitem[Liu et~al\mbox{.}(2023a)]%
        {liu2023gap}
\bibfield{author}{\bibinfo{person}{Qingyuan Liu}, \bibinfo{person}{Dong Du}, \bibinfo{person}{Yubin Xia}, \bibinfo{person}{Ping Zhang}, {and} \bibinfo{person}{Haibo Chen}.} \bibinfo{year}{2023}\natexlab{a}.
\newblock \showarticletitle{The Gap Between Serverless Research and Real-world Systems}. In \bibinfo{booktitle}{\emph{Proceedings of the 2023 ACM Symposium on Cloud Computing (Santa Cruz, CA, USA)(SoCC’23). Association for Computing Machinery, New York, NY, USA}}. \bibinfo{pages}{475--485}.
\newblock


\bibitem[Liu and Yu(2018)]%
        {liu2018elasticity}
\bibfield{author}{\bibinfo{person}{Qixiao Liu} {and} \bibinfo{person}{Zhibin Yu}.} \bibinfo{year}{2018}\natexlab{}.
\newblock \showarticletitle{The elasticity and plasticity in semi-containerized co-locating cloud workload: a view from alibaba trace}. In \bibinfo{booktitle}{\emph{Proceedings of the ACM Symposium on Cloud Computing}}. \bibinfo{pages}{347--360}.
\newblock


\bibitem[Mahgoub et~al\mbox{.}(2021)]%
        {mahgoub2021sonic}
\bibfield{author}{\bibinfo{person}{Ashraf Mahgoub}, \bibinfo{person}{Li Wang}, \bibinfo{person}{Karthick Shankar}, \bibinfo{person}{Yiming Zhang}, \bibinfo{person}{Huangshi Tian}, \bibinfo{person}{Subrata Mitra}, \bibinfo{person}{Yuxing Peng}, \bibinfo{person}{Hongqi Wang}, \bibinfo{person}{Ana Klimovic}, \bibinfo{person}{Haoran Yang}, {et~al\mbox{.}}} \bibinfo{year}{2021}\natexlab{}.
\newblock \showarticletitle{SONIC: Application-aware data passing for chained serverless applications}. In \bibinfo{booktitle}{\emph{2021 USENIX Annual Technical Conference (USENIX ATC 21)}}. \bibinfo{pages}{285--301}.
\newblock


\bibitem[Mahgoub et~al\mbox{.}(2022)]%
        {mahgoub2022orion}
\bibfield{author}{\bibinfo{person}{Ashraf Mahgoub}, \bibinfo{person}{Edgardo~Barsallo Yi}, \bibinfo{person}{Karthick Shankar}, \bibinfo{person}{Sameh Elnikety}, \bibinfo{person}{Somali Chaterji}, {and} \bibinfo{person}{Saurabh Bagchi}.} \bibinfo{year}{2022}\natexlab{}.
\newblock \showarticletitle{ORION and the three rights: Sizing, bundling, and prewarming for serverless DAGs}. In \bibinfo{booktitle}{\emph{16th USENIX Symposium on Operating Systems Design and Implementation (OSDI 22)}}. \bibinfo{pages}{303--320}.
\newblock


\bibitem[Oakes et~al\mbox{.}(2018)]%
        {oakes2018sock}
\bibfield{author}{\bibinfo{person}{Edward Oakes}, \bibinfo{person}{Leon Yang}, \bibinfo{person}{Dennis Zhou}, \bibinfo{person}{Kevin Houck}, \bibinfo{person}{Tyler Harter}, \bibinfo{person}{Andrea Arpaci-Dusseau}, {and} \bibinfo{person}{Remzi Arpaci-Dusseau}.} \bibinfo{year}{2018}\natexlab{}.
\newblock \showarticletitle{SOCK: Rapid task provisioning with Serverless-Optimized containers}. In \bibinfo{booktitle}{\emph{2018 USENIX annual technical conference (USENIX ATC 18)}}. \bibinfo{pages}{57--70}.
\newblock


\bibitem[Reiss et~al\mbox{.}(2012a)]%
        {clusterdata:Reiss2012b}
\bibfield{author}{\bibinfo{person}{Charles Reiss}, \bibinfo{person}{Alexey Tumanov}, \bibinfo{person}{Gregory~R. Ganger}, \bibinfo{person}{Randy~H. Katz}, {and} \bibinfo{person}{Michael~A. Kozuch}.} \bibinfo{year}{2012}\natexlab{a}.
\newblock \showarticletitle{Heterogeneity and dynamicity of clouds at scale: Google trace analysis}. In \bibinfo{booktitle}{\emph{ACM Symposium on Cloud Computing (SoCC)}}. \bibinfo{address}{San Jose, CA, USA}.
\newblock
\urldef\tempurl%
\url{http://www.pdl.cmu.edu/PDL-FTP/CloudComputing/googletrace-socc2012.pdf}
\showURL{%
\tempurl}


\bibitem[Reiss et~al\mbox{.}(2012b)]%
        {clusterdata:Reiss2012}
\bibfield{author}{\bibinfo{person}{Charles Reiss}, \bibinfo{person}{John Wilkes}, {and} \bibinfo{person}{Joseph~L. Hellerstein}.} \bibinfo{year}{2012}\natexlab{b}.
\newblock \showarticletitle{Obfuscatory obscanturism: making workload traces of commercially-sensitive systems safe to release}. In \bibinfo{booktitle}{\emph{3rd International Workshop on Cloud Management (CLOUDMAN)}}. \bibinfo{publisher}{IEEE}, \bibinfo{address}{Maui, HI, USA}, \bibinfo{pages}{1279--1286}.
\newblock
\urldef\tempurl%
\url{http://ieeexplore.ieee.org/xpls/abs_all.jsp?arnumber=6212064}
\showURL{%
\tempurl}


\bibitem[Sahraei et~al\mbox{.}(2023)]%
        {XFaaS_meta_sosp_23}
\bibfield{author}{\bibinfo{person}{Alireza Sahraei}, \bibinfo{person}{Soteris Demetriou}, \bibinfo{person}{Amirali Sobhgol}, \bibinfo{person}{Haoran Zhang}, \bibinfo{person}{Abhigna Nagaraja}, \bibinfo{person}{Neeraj Pathak}, \bibinfo{person}{Girish Joshi}, \bibinfo{person}{Carla Souza}, \bibinfo{person}{Bo Huang}, \bibinfo{person}{Wyatt Cook}, \bibinfo{person}{Andrii Golovei}, \bibinfo{person}{Pradeep Venkat}, \bibinfo{person}{Andrew Mcfague}, \bibinfo{person}{Dimitrios Skarlatos}, \bibinfo{person}{Vipul Patel}, \bibinfo{person}{Ravinder Thind}, \bibinfo{person}{Ernesto Gonzalez}, \bibinfo{person}{Yun Jin}, {and} \bibinfo{person}{Chunqiang Tang}.} \bibinfo{year}{2023}\natexlab{}.
\newblock \showarticletitle{XFaaS: Hyperscale and Low Cost Serverless Functions at Meta}. In \bibinfo{booktitle}{\emph{Proceedings of the 29th Symposium on Operating Systems Principles}} (Koblenz, Germany) \emph{(\bibinfo{series}{SOSP '23})}. \bibinfo{publisher}{Association for Computing Machinery}, \bibinfo{address}{New York, NY, USA}, \bibinfo{pages}{231–246}.
\newblock
\showISBNx{9798400702297}
\urldef\tempurl%
\url{https://doi.org/10.1145/3600006.3613155}
\showDOI{\tempurl}


\bibitem[Sebastio et~al\mbox{.}(2018)]%
        {clusterdata:Sebastio2018}
\bibfield{author}{\bibinfo{person}{Stefano Sebastio}, \bibinfo{person}{Kishor~S. Trivedi}, {and} \bibinfo{person}{Javier Alonso}.} \bibinfo{year}{2018}\natexlab{}.
\newblock \showarticletitle{Characterizing machines lifecycle in Google data centers}.
\newblock \bibinfo{journal}{\emph{Performance Evaluation}}  \bibinfo{volume}{126} (\bibinfo{year}{2018}), \bibinfo{pages}{39 -- 63}.
\newblock
\showISSN{0166-5316}
\urldef\tempurl%
\url{https://doi.org/10.1016/j.peva.2018.08.001}
\showDOI{\tempurl}


\bibitem[Shahrad et~al\mbox{.}(2020)]%
        {shahrad2020serverless}
\bibfield{author}{\bibinfo{person}{Mohammad Shahrad}, \bibinfo{person}{Rodrigo Fonseca}, \bibinfo{person}{Inigo Goiri}, \bibinfo{person}{Gohar Chaudhry}, \bibinfo{person}{Paul Batum}, \bibinfo{person}{Jason Cooke}, \bibinfo{person}{Eduardo Laureano}, \bibinfo{person}{Colby Tresness}, \bibinfo{person}{Mark Russinovich}, {and} \bibinfo{person}{Ricardo Bianchini}.} \bibinfo{year}{2020}\natexlab{}.
\newblock \showarticletitle{Serverless in the Wild: Characterizing and Optimizing the Serverless Workload at a Large Cloud Provider}. In \bibinfo{booktitle}{\emph{2020 USENIX Annual Technical Conference (USENIX ATC 20)}}. \bibinfo{publisher}{USENIX Association}, \bibinfo{pages}{205--218}.
\newblock
\showISBNx{978-1-939133-14-4}


\bibitem[Shi et~al\mbox{.}(2022)]%
        {shi2022characterizing}
\bibfield{author}{\bibinfo{person}{Jiuchen Shi}, \bibinfo{person}{Kaihua Fu}, \bibinfo{person}{Quan Chen}, \bibinfo{person}{Changpeng Yang}, \bibinfo{person}{Pengfei Huang}, \bibinfo{person}{Mosong Zhou}, \bibinfo{person}{Jieru Zhao}, \bibinfo{person}{Chen Chen}, {and} \bibinfo{person}{Minyi Guo}.} \bibinfo{year}{2022}\natexlab{}.
\newblock \showarticletitle{Characterizing and orchestrating VM reservation in geo-distributed clouds to improve the resource efficiency}. In \bibinfo{booktitle}{\emph{Proceedings of the 13th Symposium on Cloud Computing}}. \bibinfo{pages}{94--109}.
\newblock


\bibitem[Tirmazi et~al\mbox{.}(2020)]%
        {clusterdata:Tirmazi2020}
\bibfield{author}{\bibinfo{person}{Muhammad Tirmazi}, \bibinfo{person}{Adam Barker}, \bibinfo{person}{Nan Deng}, \bibinfo{person}{Md~E. Haque}, \bibinfo{person}{Zhijing~Gene Qin}, \bibinfo{person}{Steven Hand}, \bibinfo{person}{Mor Harchol-Balter}, {and} \bibinfo{person}{John Wilkes}.} \bibinfo{year}{2020}\natexlab{}.
\newblock \showarticletitle{Borg: the Next Generation}. In \bibinfo{booktitle}{\emph{Proceedings of the Fifteenth European Conference on Computer Systems (EuroSys'20)}}. \bibinfo{publisher}{ACM}, \bibinfo{address}{Heraklion, Greece}, Article \bibinfo{articleno}{30}, \bibinfo{numpages}{14}~pages.
\newblock
\showISBNx{9781450368827}
\urldef\tempurl%
\url{https://doi.org/10.1145/3342195.3387517}
\showDOI{\tempurl}


\bibitem[Verma et~al\mbox{.}(2015)]%
        {verma2015large}
\bibfield{author}{\bibinfo{person}{Abhishek Verma}, \bibinfo{person}{Luis Pedrosa}, \bibinfo{person}{Madhukar Korupolu}, \bibinfo{person}{David Oppenheimer}, \bibinfo{person}{Eric Tune}, {and} \bibinfo{person}{John Wilkes}.} \bibinfo{year}{2015}\natexlab{}.
\newblock \showarticletitle{Large-scale cluster management at Google with Borg}. In \bibinfo{booktitle}{\emph{Proceedings of the tenth european conference on computer systems}}. \bibinfo{pages}{1--17}.
\newblock


\bibitem[Wang et~al\mbox{.}(2021)]%
        {FaaSNet_2021}
\bibfield{author}{\bibinfo{person}{Ao Wang}, \bibinfo{person}{Shuai Chang}, \bibinfo{person}{Huangshi Tian}, \bibinfo{person}{Hongqi Wang}, \bibinfo{person}{Haoran Yang}, \bibinfo{person}{Huiba Li}, \bibinfo{person}{Rui Du}, {and} \bibinfo{person}{Yue Cheng}.} \bibinfo{year}{2021}\natexlab{}.
\newblock \showarticletitle{FaaSNet: Scalable and Fast Provisioning of Custom Serverless Container Runtimes at Alibaba Cloud Function Compute}. In \bibinfo{booktitle}{\emph{2021 USENIX Annual Technical Conference (USENIX ATC 21)}}. \bibinfo{publisher}{USENIX Association}, \bibinfo{pages}{443--457}.
\newblock
\showISBNx{978-1-939133-23-6}
\urldef\tempurl%
\url{https://www.usenix.org/conference/atc21/presentation/wang-ao}
\showURL{%
\tempurl}


\bibitem[Wang et~al\mbox{.}(2018)]%
        {wang2018}
\bibfield{author}{\bibinfo{person}{Liang Wang}, \bibinfo{person}{Mengyuan Li}, \bibinfo{person}{Yinqian Zhang}, \bibinfo{person}{Thomas Ristenpart}, {and} \bibinfo{person}{Michael Swift}.} \bibinfo{year}{2018}\natexlab{}.
\newblock \showarticletitle{Peeking Behind the Curtains of Serverless Platforms}. In \bibinfo{booktitle}{\emph{2018 USENIX Annual Technical Conference (USENIX ATC 18)}}. \bibinfo{publisher}{USENIX Association}, \bibinfo{address}{Boston, MA}, \bibinfo{pages}{133--146}.
\newblock
\showISBNx{ISBN 978-1-939133-01-4}
\urldef\tempurl%
\url{https://www.usenix.org/conference/atc18/presentation/wang-liang}
\showURL{%
\tempurl}


\bibitem[Wei et~al\mbox{.}(2023)]%
        {wei2023no}
\bibfield{author}{\bibinfo{person}{Xingda Wei}, \bibinfo{person}{Fangming Lu}, \bibinfo{person}{Tianxia Wang}, \bibinfo{person}{Jinyu Gu}, \bibinfo{person}{Yuhan Yang}, \bibinfo{person}{Rong Chen}, {and} \bibinfo{person}{Haibo Chen}.} \bibinfo{year}{2023}\natexlab{}.
\newblock \showarticletitle{No Provisioned Concurrency: Fast RDMA-codesigned Remote Fork for Serverless Computing}. In \bibinfo{booktitle}{\emph{17th USENIX Symposium on Operating Systems Design and Implementation (OSDI 23)}}. \bibinfo{pages}{497--517}.
\newblock


\bibitem[Weng et~al\mbox{.}(2022)]%
        {weng2022mlaas}
\bibfield{author}{\bibinfo{person}{Qizhen Weng}, \bibinfo{person}{Wencong Xiao}, \bibinfo{person}{Yinghao Yu}, \bibinfo{person}{Wei Wang}, \bibinfo{person}{Cheng Wang}, \bibinfo{person}{Jian He}, \bibinfo{person}{Yong Li}, \bibinfo{person}{Liping Zhang}, \bibinfo{person}{Wei Lin}, {and} \bibinfo{person}{Yu Ding}.} \bibinfo{year}{2022}\natexlab{}.
\newblock \showarticletitle{MLaaS in the wild: Workload analysis and scheduling in Large-Scale heterogeneous GPU clusters}. In \bibinfo{booktitle}{\emph{19th USENIX Symposium on Networked Systems Design and Implementation (NSDI 22)}}. \bibinfo{pages}{945--960}.
\newblock


\bibitem[Weng et~al\mbox{.}(2023)]%
        {weng2023beware}
\bibfield{author}{\bibinfo{person}{Qizhen Weng}, \bibinfo{person}{Lingyun Yang}, \bibinfo{person}{Yinghao Yu}, \bibinfo{person}{Wei Wang}, \bibinfo{person}{Xiaochuan Tang}, \bibinfo{person}{Guodong Yang}, {and} \bibinfo{person}{Liping Zhang}.} \bibinfo{year}{2023}\natexlab{}.
\newblock \showarticletitle{Beware of Fragmentation: Scheduling GPU-Sharing Workloads with Fragmentation Gradient Descent}. In \bibinfo{booktitle}{\emph{2023 USENIX Annual Technical Conference (USENIX ATC 23)}}. \bibinfo{pages}{995--1008}.
\newblock


\bibitem[Wilkes(2020)]%
        {clusterdata:Wilkes2020a}
\bibfield{author}{\bibinfo{person}{John Wilkes}.} \bibinfo{year}{2020}\natexlab{}.
\newblock \bibinfo{booktitle}{\emph{Google cluster-usage traces v3}}.
\newblock \bibinfo{type}{Technical Report}. \bibinfo{institution}{Google Inc.}, \bibinfo{address}{Mountain View, CA, USA}.
\newblock
\newblock
\shownote{Posted at \url{https://github.com/google/cluster-data/blob/master/ClusterData2019.md}}.


\bibitem[Zeng et~al\mbox{.}(2022)]%
        {zeng2022congestion}
\bibfield{author}{\bibinfo{person}{Gaoxiong Zeng}, \bibinfo{person}{Wei Bai}, \bibinfo{person}{Ge Chen}, \bibinfo{person}{Kai Chen}, \bibinfo{person}{Dongsu Han}, \bibinfo{person}{Yibo Zhu}, {and} \bibinfo{person}{Lei Cui}.} \bibinfo{year}{2022}\natexlab{}.
\newblock \showarticletitle{Congestion control for cross-datacenter networks}.
\newblock \bibinfo{journal}{\emph{IEEE/ACM Transactions on Networking}} \bibinfo{volume}{30}, \bibinfo{number}{5} (\bibinfo{year}{2022}), \bibinfo{pages}{2074--2089}.
\newblock


\bibitem[Zhang et~al\mbox{.}(2021)]%
        {zhang2021faster}
\bibfield{author}{\bibinfo{person}{Yanqi Zhang}, \bibinfo{person}{Inigo Goiri}, \bibinfo{person}{Gohar~Irfan Chaudhry}, \bibinfo{person}{Rodrigo Fonseca}, \bibinfo{person}{Sameh Elnikety}, \bibinfo{person}{Christina Delimitrou}, {and} \bibinfo{person}{Ricardo Bianchini}.} \bibinfo{year}{2021}\natexlab{}.
\newblock \showarticletitle{Faster and cheaper serverless computing on harvested resources}. In \bibinfo{booktitle}{\emph{Proceedings of the ACM SIGOPS 28th Symposium on Operating Systems Principles}}. \bibinfo{pages}{724--739}.
\newblock


\bibitem[Zhuang et~al\mbox{.}(2023)]%
        {zhuang2023exoflow}
\bibfield{author}{\bibinfo{person}{Siyuan Zhuang}, \bibinfo{person}{Stephanie Wang}, \bibinfo{person}{Eric Liang}, \bibinfo{person}{Yi Cheng}, {and} \bibinfo{person}{Ion Stoica}.} \bibinfo{year}{2023}\natexlab{}.
\newblock \showarticletitle{ExoFlow: A universal workflow system for Exactly-Once DAGs}. In \bibinfo{booktitle}{\emph{17th USENIX Symposium on Operating Systems Design and Implementation (OSDI 23)}}. \bibinfo{pages}{269--286}.
\newblock


\end{thebibliography}

\end{document}